\documentclass[aps,prb,twocolumn,amsmath,amssymb,nofootinbib,eqsecnum,tighten,showpacs,floatfix]{revtex4}

\usepackage[dvips]{color} 
\usepackage{graphicx}
\usepackage{dcolumn} 
\usepackage{bm}      
\usepackage{epic}
\usepackage{longtable}
\usepackage{curves}

\begin{document}

\title{
Crossover of conductance and local density of states
in a single-channel disordered quantum wire
      }

\author{S.\ Ryu}
\affiliation{Department of Applied Physics,
             University of Tokyo,
             7-3-1 Hongo Bunkyo-ku,
             Tokyo 113-8656,
             Japan}
\author{C.\ Mudry}
\affiliation{Paul Scherrer Institute,
             CH-5232 Villigen PSI, 
             Switzerland}
\author{A.\ Furusaki}
\affiliation{Condensed-Matter Theory Laboratory, 
             RIKEN, 
             Wako, 
             Saitama 351-0198, 
             Japan}
\date{\today}

\begin{abstract}
The probability distribution of the mesoscopic local density of states
(LDOS) for a single-channel disordered quantum wire with chiral symmetry is
computed in two different geometries.
An approximate ansatz is proposed to describe the crossover of the probability 
distributions for the conductance and LDOS between the chiral and standard 
symmetry classes of a single-channel disordered quantum wire.
The accuracy of this ansatz is discussed by comparison with
a large-deviation ansatz introduced by H.\ Schomerus and M.\ Titov
in Phys.\ Rev.\ B \textbf{67}, 100201(R) (2003).
\end{abstract}

\maketitle

\section{
Introduction
        }
\label{Introduction}

The metal-insulator transition induced by disorder in the problem of
Anderson localization remains poorly understood from a 
theoretical point of view. 
On the one hand, it is true that the prediction 
for the existence of a metal-insulator transition induced by disorder
has been verified by numerical simulations.
On the other hand, the critical exponents at the metal-insulating transition
measured numerically still cannot be extracted from the field theories which
are believed to embody the metal-insulator transition.\cite{Kramer93} 
The paradigm of this
unpleasant situation is the plateau transition in the lowest Landau level of 
the integer quantum Hall effect.\cite{Huckestein95}  
The development of
reliable methods to compute analytically critical properties at a 
disorder-induced metal-insulator transition remains to this date an open problem.

The simplest known example of disorder-induced criticality
in the context of Anderson localization
was solved by Dyson in 1953.\cite{Dyson53}
According to Dyson, the global density of states (DOS) 
in the thermodynamic limit and close to the band center is anomalous
when a single quantum particle hops with a random amplitude
between the nearest-neighbor sites of a chain
excluding any other form of disorder, 
in short, the one-dimensional random hopping problem.
Subsequent works 
showed that Dyson's spectral anomaly is related to 
a diverging localization length,\cite{Theodorou76,Eggarter78,Balents97}
an anomalous decay of the envelope of wave functions,\cite{Fleishman77} 
and an anomalous probability distribution of the resistance 
upon approaching the band center.\cite{Stone81}

The coalescence of the concept of universality 
imported from the theory of critical phenomena
with the symmetry classification imported 
from random matrix theory has led to
a classification of diffusive regimes and 
disorder-induced metal-insulator transitions
in terms of 10 universality classes that are 
uniquely characterized by the 
intrinsic symmetries preserved by the disorder (assumed weak)
for any given dimensionality of space.\cite{Zirnbauer96}
According to this classification, Dyson's singular 
DOS is a trademark of the chiral universality classes.
The chiral universality classes are realized in problems of Anderson
localization for which the Hamiltonian anticommutes with some unitary
operator for all realizations of the disorder and in which case the DOS
is known to be singular at the band center in 
zero,\cite{chRMT,Fyodorov03}
quasi-one,\cite{Brouwer00,Titov01,Altland01}
and two dimensions.\cite{Gade93,Motrunich02,Horovitz02,Mudry03}
Even though the chiral universality classes are 
critical at the band center,
they have
been far less studied than the orthogonal, unitary, and symplectic
universality classes, which we will refer to 
as the standard universality classes.
For example, 
the probability distribution of the so-called mesoscopic 
local density of states (LDOS) has been known since 1989 from the work of
Altshuler and Prigodin for a single chain with weak on-site 
disorder,\cite{Altshuler91,Bunder01,Schomerus02}
whereas it has not yet been computed for the 
one-dimensional random-hopping problem.
The first aim of this paper is to fill this gap.

If a diffusive regime exists, perturbative techniques can be used
to describe the crossovers 
between universality classes.\cite{Mudry00,Brouwer03}
However, little is known quantitatively
on the crossover between two different universality classes,
here defined in terms of nonlinear-sigma models, say,
from the diffusive to the localized regimes.
This is even true in one dimension, which is by far the most studied
laboratory for Anderson localization as was recently illustrated by a flurry
of works on the validity of one-parameter scaling 
in one dimension.\cite{Deych98-01,Schomerus03a,Schomerus03b,Titov03,Dossetti-Romero04}
The absence of a diffusive regime renders the concept of
universality classes ambiguous in one dimension
in that  symmetry alone does not
specify a universality class in one dimension.
For example,
the ratio of the localization length to the mean free path
in a wire of finite width 
interpolates smoothly between its two limiting values for 
fully preserved 
or 
completely broken
time-reversal symmetry, respectivley, as a function of
a weak magnetic field.\cite{Schomerus00}
Similarly, 
fine tuning of microscopic parameters is required
to achieve delocalization (i.e., diverging localization length)
of quasiparticles at the fermi energy
in a dirty superconducting wire of finite thickness 
with both broken spin-rotation and time-reversal symmetries
whereas delocalization becomes generic 
in the thick wire scaling limit.\cite{Brouwer03,Motrunich01}
One can nevertheless define a
\textit{symmetry class} in one dimension
by demanding that single-parameter scaling holds.
Lack of universality in one dimension can then be understood
as the fact that there is no scaling limit for which
single-parameter scaling becomes a generic
property of an ensemble of random microscopic Hamiltonians of a given
symmetry. Symmetry classes in one dimension
cannot be construed as an enumeration of stable or unstable
fixed points of some putative effective field theory.
Computing the crossover between symmetry classes in one dimension
nevertheless remains a well-defined problem.

Time-reversal symmetry cannot be broken for a spinless particle
constrained to move in a one-dimensional and simply connected world,
in which case the only possible crossover between symmetry classes
takes place 
between the chiral and standard classes.
The second aim of this paper is to compute the crossover 
between these two symmetry classes in one dimension 
for the probability distributions of the conductance and of the LDOS
from the ballistic to the localized regimes 
assuming a weak disorder at the microscopic level. 
In that regard, our results are complementary to the analysis made
by Schomerus and Titov of the validity of one-parameter scaling
in one dimension in which they computed exactly the first four cumulants 
of the logarithm of the conductance to leading order in the ratio of the
length of the chain to the localization length
assuming a weak disorder at the microscopic
level.\cite{Schomerus03a,Schomerus03b}   
We thus propose an approximate ansatz for the crossover whereby 
the approximation consists of assuming that the phase and amplitude
of the reflection coefficient separate. By comparing the cumulants
of the Lyapunov exponent of the transfer matrix computed with our
approximate ansatz to the ones computed with the large-deviation 
ansatz of Schomerus and Titov deep in the localization regime, 
we deduce that our ansatz captures the first cumulant, but fails with
the higher cumulants.

The paper is organized as follows.
We formulate the one-dimensional problem of Anderson localization 
as a continuous Schr\"odinger equation with a relativistic
kinetic energy
in 
Sec.\  \ref{Model}.
A functional renormalization group equation is derived
for the probability distribution of the reflection coefficient
in Sec.\  \ref{Functional renormalization group equations}.
Exact solutions to these 
coupled functional renormalization group equations
are given for the chiral and the standard classes,
as well as an approximate solution that describes the
crossovers between these two classes
in Sec.\  
\ref{sec: Solutions to the functional renormalization group equations}.
The accuracy of the approximate solution for the crossover regime
is estimated in
Sec.\  
\ref{sec: Validity of the separation ansatz for the crossover regime}.
The probability distribution of the LDOS is computed exactly in
the chiral and standard classes 
whereas it is computed approximately in the crossover regime between 
these two classes in
Sec.\  \ref{sec: Probability distribution of the LDOS}.
We conclude in Sec.\ \ref{sec: Conclusion}.

\begin{figure}[t] 
\begin{center}
\begin{picture}(220,150)(-120,-130)

\thicklines
\put(-80, 20){\line(1,0){140}}
\put(-80,-10){\line(1,0){140}}
\put(60,-10){\line(0,1){30}}

\thinlines

\put( -40,  5){\arc   ( 0,10){-180}}
\put( -40, 15){\line  (-1, 0){  20}}
\put( -40,-5){\vector(-1, 0){  20}}
\put( -120,  17){$\mathrm{(a)}$}
\put( -95,  5){$r$}
\put( -110,  -5){$(|r|=1)$}

\put( -70,-10){\line(0,1){30}}

\multiput(-70,20)(10,0){10}{\line(1,-1){30}}

\put(-70,10){\line(1,-1){20}}
\put(-70,0){\line(1,-1){10}}

\put(30,20){\line(1,-1){30}}
\put(40,20){\line(1,-1){20}}
\put(50,20){\line(1,-1){10}}

\thicklines
\put(-80,-30){\line(1,0){140}}
\put(-80,-60){\line(1,0){140}}
\put(60,-60){\line(0,1){30}}

\thinlines

\put( -40,-45){\arc   ( 0,10){-180}}
\put( -40,-35){\line  (-1, 0){  20}}
\put( -40,-55){\vector(-1, 0){  20}}
\put( -120,  -33){$\mathrm{(b)}$}
\put( -95,-45){$r$}

\put( -70,-60){\line(0,1){30}}

\multiput(-70,-30)(10,0){10}{\line(1,-1){30}}

\put(-70,-40){\line(1,-1){20}}
\put(-70,-50){\line(1,-1){10}}

\put(30,-30){\line(1,-1){30}}
\put(40,-30){\line(1,-1){20}}
\put(50,-30){\line(1,-1){10}}

\thicklines
\put(-80, -80){\line(1,0){150}}
\put(-80,-110){\line(1,0){150}}

\thinlines

\put( -40,  -95){\arc   ( 0,10){-180}}
\put( -40, -85){\line  (-1, 0){  20}}
\put( -40,-105){\vector(-1, 0){  20}}
\put( -120,  -83){$\mathrm{(c)}$}
\put( -95,  -95){$r$}

\put( -70,-110){\line(0,1){30}}
\put(60,-110){\line(0,1){30}}

\multiput(-70,-80)(10,0){10}{\line(1,-1){30}}

\put(-70,-90){\line(1,-1){20}}
\put(-70,-100){\line(1,-1){10}}

\put(30,-80){\line(1,-1){30}}
\put(40,-80){\line(1,-1){20}}
\put(50,-80){\line(1,-1){10}}

\put(-70,-120){\vector(1,0){130}}
\put(60,-120){\vector(-1,0){130}}
\put(-5, -130){$L$}

\end{picture}
\end{center}
\caption{
\label{fig: geometries}
In this paper,
we will consider two different simply-connected geometries
for a strictly 1D disordered quantum wire which are
imposed by suitable boundary conditions at $y=-L/2$ and $y=+L/2$, 
respectively. In geometry (a) and (b) the disordered quantum wire
is closed to the right and open to the left where it is
connected to an ideal lead. The particle in geometry (b) is
subjected to absorption whereas it is not in geometry (a).
In geometry (c) the disordered quantum wire
is open to the left and to the right where it is
connected to ideal leads.
        }
\end{figure}
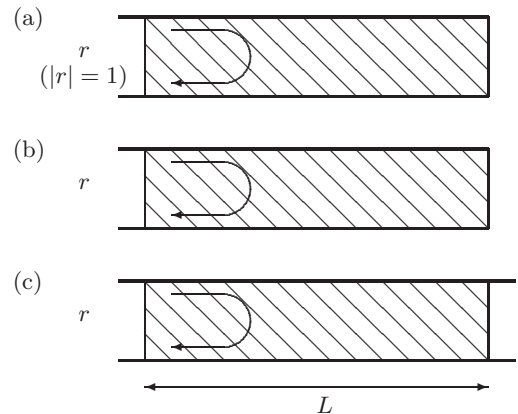

\section{
Model
        }
\label{Model}

In this paper we want to investigate the statistical properties
of the LDOS $\nu$ and dimensionless conductance $g$ 
for a single quantum particle, which is (i) restricted to a simply connected 
and strictly one-dimensional (1D) geometry 
and (ii) subjected to a static and weak random environment 
(disorder) that interpolates smoothly between the 
standard and chiral symmetry classes. 
Condition (i) implies that
time-reversal symmetry is preserved for every realization of the disorder.
In 1D and assuming a metallic ground state in the absence of the disorder,
the ground state consists of two distinct Fermi points.
Plane waves at the Fermi points are called left- and right-movers, 
respectively. 
In the standard symmetry class,
the channels for forward and backward scatterings induced 
by an on-site disorder potential in the basis of left- and right-movers
are equally likely up to nonrandom oscillatory factors. 
In the chiral symmetry class, 
the disorder potential is off-diagonal
in the basis of left- and right-movers.
Presuming weak disorder in
condition (ii) allows us to choose
a kinetic energy that is a first-order differential operator, for
it only makes sense to linearize the spectrum about the two
Fermi points when the disorder potential is weak.
We thus choose to model the dynamics of the single quantum particle
and the static random environment by the Hamiltonian
\begin{subequations}
\label{eq: def our Anderson localization problem} 
\begin{eqnarray}
H:=
-
\sigma_{3}
{i}
\frac{{d}}{{d} y}
-
\sum_{\mu=0}^2\sigma^{\ }_{\mu}v^{\ }_{\mu}(y),
\label{eq: def our Anderson localization problem a} 
\end{eqnarray}
where we have set the Fermi velocity 
$v^{\ }_{\mathrm{F}}$
and
$\hbar$
to one.
The boundary conditions that will be imposed in the sequel are depicted 
in Fig.\ \ref{fig: geometries}. 
They all obey condition (i). We have denoted by
$\sigma^{\ }_{0}$ the unit $2\times 2$ matrix and by  
$\sigma^{\ }_{1,2,3}$ the $2\times2$ Pauli matrices.
The static random environment is represented by three independent
potentials $v_{\mu}(y)\in \mathbb{R}$, $\mu=0,1,2$ that we choose,
for mathematical convenience, 
to be supported on the interval $-L/2\leq y\leq+L/2$,
and to be white-noise and Gaussian distributed with vanishing mean,
(disorder averaging is denoted by $\langle \ldots \rangle$)
\begin{eqnarray}
&&
\left\langle
v^{\ }_{\mu}(y)
\right\rangle
=0,
\label{eq: def our Anderson localization problem b} 
\\
&&
\left\langle
v^{\ }_{\mu}(y)
v^{\ }_{\nu}(y^{\prime})
\right\rangle=
2
g^{\ }_{\mu}
\delta^{\ }_{\mu\nu}
\delta(y-y^{\prime}).
\label{eq: def our Anderson localization problem c} 
\end{eqnarray}
\end{subequations}
The three variances $g^{\ }_{0,1,2}$ carry the dimensions of inverse length
and need not be equal. 
A finite value 
\begin{eqnarray}
\left\langle v^{\ }_2(y)\right\rangle=\Delta
\end{eqnarray}
can also be easily accommodated in our formalism.\cite{Brouwer98}
In this case, the clean system is insulating as 
$\Delta$ opens up a dimerization gap at the Fermi energy
and the random environment mimics to a first approximation the static
fluctuations of the phonons responsible for the dimerization
sufficiently close to the Peierls transition.\cite{Lee73}
The problem of Anderson localization in the continuum
defined by Eq.\ (\ref{eq: def our Anderson localization problem})
is a coarse-grained version of a single particle hopping on a 
simply-connected chain with a uniform nearest-neighbor hopping amplitude
subjected to weak and independent random fluctuations of an on-site potential 
and of the nearest-neighbor hopping amplitude.

The generic symmetry obeyed by
Eq.\ (\ref{eq: def our Anderson localization problem a})
is time-reversal invariance,
\begin{eqnarray}
\sigma^{\ }_1\, H^*\, \sigma^{\ }_1= \hphantom{+}H.
\label{eq: def TR symmetry}
\end{eqnarray}
If the condition
\begin{eqnarray}
g^{\ }_0=
g^{\ }_1=
0
\label{eq: def chiral limit for g's}
\end{eqnarray}
is satisfied, or, for that matter, 
any equivalent condition obtained by a redefinition
of the Pauli matrices through
a position independent SU(2) rotation around $\sigma^{\ }_3$,
Eq.\ (\ref{eq: def our Anderson localization problem a})
has an additional chiral symmetry in that
it anticommutes with $\sigma^{\ }_1$,
\begin{eqnarray}
\sigma^{\ }_1\, H\, \sigma^{\ }_1= -H.
\label{eq: def chiral symmetry for H}
\end{eqnarray}
Consequently, all nonvanishing eigenvalues of 
$H$ come in pairs with opposite signs.
The origin of the chiral symmetry of the continuum model 
for any given realization of the disorder
is a sublattice symmetry of an appropriate lattice regularization.
A microscopic random Hamiltonian realizing the chiral symmetry
is, for example, a tight-binding Hamiltonian on a hypercubic lattice
with strictly vanishing on-site energies
and random hopping matrix elements restricted to
nearest-neighbor sites, in short, the random hopping problem.

The LDOS, DOS, conductance, shot-noise power, etc.,
can all be extracted from the elements of 
the $2\times2$ scattering matrix 
$S^{\ }_E$
at the energy $E$
that connects incoming (``i'') and outgoing (``o'') states
\begin{eqnarray}
\left(
\begin{array}{c}
\psi^{\mathrm{o,L}} \\
\psi^{\mathrm{o,R}}
\end{array}
\right)^{\ }_{E}
&=&
S^{\ }_{E}
\left(
\begin{array}{c}
\psi^{\mathrm{i,L}} \\
\psi^{\mathrm{i,R}}
\end{array}
\right)^{\ }_{E}
\nonumber\\
&\equiv&
\left(
\begin{array}{cc}
r & t^{\prime} \\
t & r^{\prime}
\end{array}
\right)^{\ }_{E}
\left(
\begin{array}{c}
\psi^{\mathrm{i,L}} \\
\psi^{\mathrm{i,R}}
\end{array}
\right)^{\ }_{E}.
\label{eq: def S matrix}
\end{eqnarray}
The amplitude for incoming and outgoing waves 
to the left- and right-hand side of
the disordered region is denoted by
$\psi^{\mathrm{i/o},\mathrm{L/R}}$
here. The matrix elements
$r, r^{\prime} \in \mathbb{C}$ 
are the reflection coefficients
whereas the matrix elements
$t, t^{\prime} \in \mathbb{C}$ are the transmission coefficients.
Alternatively, one can work with the $2\times2$ transfer matrix
$\mathcal{M}^{\ }_E$
at the energy $E$ 
which follows from the scattering matrix (\ref{eq: def S matrix}) 
through the basis transformation implied by
\begin{eqnarray}
\left(\begin{array}{c}
\psi^{\mathrm{o,R}}\\
\psi^{\mathrm{i,R}}\\
\end{array}\right)_{E}
=
\mathcal{M}_{E}
\left(\begin{array}{c}
\psi^{\mathrm{i,L}}\\
\psi^{\mathrm{o,L}}\\
\end{array}\right)_{E}.
\label{eq: def M matrix}
\end{eqnarray}
Conservation of probability dictates that 
the scattering matrix is unitary or, equivalently, that 
the transfer matrix is pseudo-unitary
\begin{subequations}
\begin{eqnarray}
&&
\left(S^{\ }_E\right)^{\dag}\,\sigma^{\ }_0\,S^{\ }_E=
S^{\ }_E\,\sigma^{\ }_0\,\left(S^{\ }_E\right)^{\dag}=
\sigma^{\ }_0,
\\
&&
\left(\mathcal{M}^{\ }_E\right)^{\dag}\,\sigma^{\ }_3\,\mathcal{M}^{\ }_E=
\mathcal{M}^{\ }_E\,\sigma^{\ }_3\,\left(\mathcal{M}^{\ }_E\right)^{\dag}=
\sigma^{\ }_3.
\end{eqnarray}
\end{subequations}
Conservation of probability
also implies the existence of the polar decompositions
\begin{subequations}
\label{eq: polar decomposition S+M}
\begin{eqnarray}
S^{\ }_{E}&=&
\left(
\begin{array}{cc}
v^{\prime *} & 0 \\
0 & u
\end{array}
\right)^{\ }_E\!
\left(
\begin{array}{cc}
-\tanh x & \mathrm{sech}\, x \\
\mathrm{sech}\, x  & \tanh x
\end{array}
\right)^{\ }_E\!
\left(
\begin{array}{cc}
v & 0 \\
0 & u^{\prime *}
\end{array}
\right)^{\ }_E,
\nonumber\\
&&\\
\mathcal{M}^{\ }_{E}&=&
\left(
\begin{array}{cc}
u & 0 \\
0 & u^{\prime}
\end{array}
\right)^{\ }_E
\left(
\begin{array}{cc}
\cosh x & \sinh x \\
\sinh x & \cosh x
\end{array}
\right)^{\ }_E
\left(
\begin{array}{cc}
v & 0 \\
0 & v^{\prime}
\end{array}
\right)^{\ }_E,
\nonumber\\
&&
\end{eqnarray}
\end{subequations}
where 
$u^{\ }_E$,
$u^{\prime}_E$,
$v^{\ }_E$,
$v^{\prime}_E$ are independent complex numbers with
$|u^{\ }_E|^{2}=|u^{\prime}_E|^{2}=|v^{\ }_E|^{2}=|v^{\prime}_E|^{2}=1$
and
$
x^{\ }_E\in\mathbb{R}
$.
The polar decomposition is not unique.
For example, the sign of $x^{\ }_E$ can be absorbed into a redefinition
of 
$u^{\ }_E$,
$u^{\prime}_E$,
$v^{\ }_E$,
$v^{\prime}_E$.
The variable $x^{\ }_E$, when restricted to the halfline
$[0,\infty[$, has the geometrical interpretation of a radial coordinate on a 
Riemannian manifold.\cite{Hueffmann90,Brouwer00-nonuni}
The time-reversal symmetry of the Hamiltonian implies the transformation laws
\begin{subequations}
\begin{eqnarray}
&&
S^{\ }_E=\left(S^{\ }_E\right)^{{T} },
\\
&&
\sigma^{\ }_1\,\mathcal{M}^{* }_{E}\,\sigma^{\ }_1=
\mathcal{M}^{\ }_{E},
\end{eqnarray}
\end{subequations}
which enforce the constraints 
$u^{\prime}_E=u^{* }_E$
and  
$v^{\prime}_E=v^{* }_E$
or,
equivalently,
$t^{\ }_E=t^{\prime}_E$.
The chiral symmetry of the Hamiltonian implies the transformation laws
\begin{subequations}
\begin{eqnarray}
&&
S^{\ }_{+E}=\left(S^{\ }_{-E}\right)^{\dag},
\label{eq: chiral symmetry S}
\\
&&
\sigma^{\ }_1\,\mathcal{M}^{\ }_{+E}\,\sigma^{\ }_1=
\mathcal{M}^{\ }_{-E},
\label{eq: chiral symmetry M}
\end{eqnarray}
\end{subequations}
which enforce the constraints 
$u^{\prime}_{+E}=u^{\ }_{-E}$
and  
$v^{\prime}_{+E}=v^{\ }_{-E}$
or,
equivalently,
$r^{\ }_{+E}=r^{* }_{-E}$,
$r^{\prime}_{+E}=r^{\prime\, *}_{-E}$,
$t^{\ }_{+E}=t^{\prime\,*}_{-E}$.
Then, at the band center $E=0$
with the condition (\ref{eq: def chiral limit for g's}), 
the scattering matrix becomes Hermitian.

Under the assumptions that
(i) the disorder is weak
and (ii)
$u^{\ }_E$,
$u^{\prime}_E$,
$v^{\ }_E$,
$v^{\prime}_E$
are all independently and uniformly distributed on the unit circle
in the complex plane,
it was essentially shown in Ref.\ \onlinecite{Anderson80} 
that the probability distribution 
$\mathcal{X}$
of 
$x^{\ }_E\geq0$
obeys the Fokker-Planck equation
\begin{eqnarray}
\frac{\partial \mathcal{X}(x;L)}{\partial L}=
\frac{1}{4\ell}
\frac{\partial  }{\partial x}
\sinh(2x)
\frac{\partial  }{\partial x}
\mathrm{csch}(2x)
\mathcal{X}(x;L)
\hphantom{AA}
\label{eq: FP std limit}
\end{eqnarray}
in the geometry of Fig.\ \ref{fig: geometries}c.
The mean free path $\ell$ is 
some function of 
$g^{\ }_{0}$, $g^{\ }_{1}$, $g^{\ }_{2}$, and $\varepsilon$.
In a more general context of multi channel quantum wires
assumption (ii) is known as the isotropy assumption,\cite{Mello88}
whereas the Fokker-Planck equation
(\ref{eq: FP std limit})
is a special case of the
Dorokhov-Mello-Pichard-Kumar (DMPK) equation.\cite{Dorokhov82,Mello88}
A consequence of assumption (ii) is one-parameter scaling as encoded by
the Fokker-Planck equation (\ref{eq: FP std limit}), i.e.,
the probability distribution of $x$ and, consequently, of 
$g=\mathrm{sech}^2\,x$
depends on the single dimensionless parameter $L/\ell$.
It is evident that assumption (ii) breaks down 
at the band center $E=0$ and
with the chiral condition
(\ref{eq: def chiral limit for g's}) 
because
$u^{\ }_{E=0}=u^{\prime}_{E=0}$ 
and
$v^{\ }_{E=0}=v^{\prime}_{E=0}$ 
are then both real valued. Correspondingly, 
the Fokker-Planck equation obeyed by the probability distribution 
$\mathcal{X}$
of
$-\infty< x^{\ }_{E=0}<+\infty$
is different from
Eq.\ (\ref{eq: FP std limit})
and given by the diffusion equation
\begin{eqnarray}
\frac{\partial \mathcal{X}(x;L)}{\partial L}=
\frac{1}{2\ell}
\frac{\partial^2  \mathcal{X}(x;L)}{\partial x^2}
\label{eq: FP chiral limit}
\end{eqnarray}
in the geometry of 
Fig.\ \ref{fig: geometries}c.\cite{Stone81,Mathur97,Brouwer98}
The mean free path $\ell$ is 
some function of 
$g^{\ }_{2}$.
Again, the Fokker-Planck equation (\ref{eq: FP chiral limit})
encodes one-parameter scaling as the probability distribution of $x$
and, consequently, of $g=\mathrm{sech}^2\,x$
depends on the single dimensionless parameter $L/\ell$.
There are no known multiple-parameter scaling equations for the 
LDOS, DOS, conductance, etc.,
which describe the crossover between 
the two limiting cases described by
Eqs.\ (\ref{eq: FP std limit})
and 
(\ref{eq: FP chiral limit})
for values of $L$ ranging from the ballistic 
($L$ smaller than the mean free path)
to the localized regime
($L$ larger than the mean free path)
to the best of our knowledge
(see Refs.\ \onlinecite{Shapiro86,Shapiro87,Cohen88}
for discussions of two-parameter scaling). 
We fill this gap in the remainder of the paper.

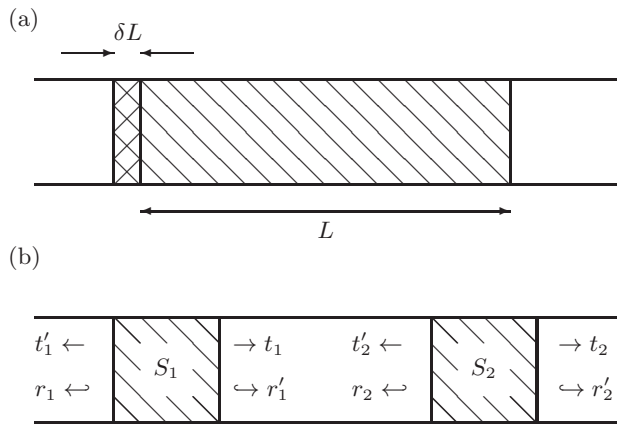
\begin{figure}[t] 
\begin{center}
\begin{picture}(240,160)(0,-140) 
\put(0,10){(a)}
\thicklines
\put(  +10,-10){\line(1,0){220}}
\put(  +10,-50){\line(1,0){220}}
%
%
\put( 50,-50){\line(0,1){40}}
\put(190,-50){\line(0,1){40}}
%
%
\thinlines
\put(120,-60){\vector(-1,0){70}}
\put(120,-60){\vector( 1,0){70}}
\put(117,-70){$L$}
\thicklines
\put( 40,-50){\line(0,1){40}}    
\thinlines
\put( 20,  0){\vector( 1,0){20}} 
\put( 70,  0){\vector(-1,0){20}} 
\put( 40,  5){$\delta L$}        
\put(50,-20){\line(1,-1){30}}
\put(50,-30){\line(1,-1){20}}
\put(50,-40){\line(1,-1){10}}
\multiput(50,-10)(10,0){11}{\line(1,-1){40}} 
\put(190,-20){\line(-1,1){10}}
\put(190,-30){\line(-1,1){20}}
\put(190,-40){\line(-1,1){30}}
%
 \multiput( 40,-50)(0,10){4}{\line(1,1){10}} 
 \multiput( 50,-50)(0,10){4}{\line(-1,1){10}} 
%
%
\put(  0,-80){(b)}
%
%
\thicklines
\put(  10,-100){\line(1,0){220}}
\put(  10,-140){\line(1,0){220}}
%
%
\multiput(40,-140)(40,0){2}{\line(0,1){40}} 
\multiput(160,-140)(40,0){2}{\line(0,1){40}} 
\thinlines
%
%
\multiput(80,-140)(-10,0){4}{\line(-1,1){10}} 
\multiput(40,-100)(10, 0){4}{\line(1,-1){10}} 
\multiput(40,-100)(0,-10){4}{\line(1,-1){10}} 
\multiput(80,-140)(0, 10){4}{\line(-1,1){10}} 
%
%
\multiput(200,-140)(-10,0){4}{\line(-1,1){10}} 
\multiput(160,-100)(10, 0){4}{\line(1,-1){10}} 
\multiput(160,-100)(0,-10){4}{\line(1,-1){10}} 
\multiput(200,-140)(0, 10){4}{\line(-1,1){10}} 
%
%
\put(  10,-113){$t'    _1\leftarrow    $}
\put(  10,-130){$r^{\ }_1\hookleftarrow$}
\put(208,-113){$\rightarrow     t^{\ }_2$}
\put(208,-130){$\hookrightarrow r'    _2$}
\put( 85,-113){$\rightarrow     t^{\ }_1$}
\put( 85,-130){$\hookrightarrow r'    _1$}
\put(130,-113){$t'    _2\leftarrow    $}
\put(130,-130){$r^{\ }_2\hookleftarrow$}
%
%
\put( 55,-122){$S_1$}
\put(175,-122){$S_2$}
%
%
\end{picture}
\end{center}
\caption{
\label{fig:composition law and RG step}
(a) A thin slice of length $\delta L$ with $a\ll\delta L\ll\ell\ll L$
is added to the left of the disordered region of length $L$. 
(b) Two disordered regions 1 and 2 with scattering matrices 
$S_1$ 
and 
$S_2$, 
respectively, in a quantum wire.
        }
\end{figure}

\section{
Functional renormalization group equations
        }
\label{Functional renormalization group equations}

We shall show in Secs.\ 
\ref{sec: Solutions to the functional renormalization group equations}
and 
\ref{sec: Probability distribution of the LDOS}
that the statistical properties of the dimensionless conductance $g$
and of the LDOS $\nu$ follow from the knowledge of the statistical properties
of the reflection coefficient 
\begin{eqnarray}
r=:
\sqrt{R}\,
\exp\left({i}\phi\right),
\label{eq: def R phi}
\end{eqnarray}
which has been decomposed into its square modulus 
$0\leq R\leq1$ and phase $0\leq\phi<2\pi$.
For a disordered wire of length
$L+\delta L$ as depicted in Fig.\ \ref{fig:composition law and RG step},
the reflection coefficient 
$r^{\ }_{L+\delta L}$ 
is related to the entries of the 
scattering matrix $S^{\ }_{\delta L}$ for the slice of length $\delta L$
and to the entry $r^{\ }_{L}$ of the scattering matrix $S^{\ }_{L}$
for the much longer segment of length $L$ by the composition law
\begin{equation}
r^{\ }_{L+\delta L} =
r^{\ }_{  \delta L}
+
t'_{  \delta L}
\left(
{1}-r^{\ }_{L        }r'_{  \delta L}
\right)^{-1} 
r^{\ }_{L        }t^{\ }_{  \delta L}.
\label{eq:composition law reflection amplitude}
\end{equation}
When the width $\delta L$ of the slice is much larger than the lattice
spacing $a$ but much smaller than the mean free path $\ell$,
here defined by
\cite{Mudry00}
\begin{eqnarray}
\left\langle
r^{\ }_{\delta L}
r^{* }_{\delta L}
\right\rangle
=:
\frac{\delta L}{\ell},
\label{eq: def ell}
\end{eqnarray}
we infer the continuous Langevin process
\begin{subequations}
\label{eq: Langevin eq for R and phi}
\begin{eqnarray}
\frac{{d}R}{{d}L}&=&
-4\mathcal{E}^{\prime\prime} R
+2\sqrt{R}(1-R)
(v_{1}\sin\phi-v_{2}\cos\phi),
\hphantom{AAA}
\label{eq: Langevin eq for R} \\
\frac{{d}\phi}{{d}L}
&=&
2(\mathcal{E}^{\prime}+v_{0})
+(v_{1}\cos\phi+v_{2}\sin\phi)
\left(
\sqrt{R}+\frac{1}{\sqrt{R}}
\right),
\nonumber\\&&
\label{eq: Langevin eq for phi} 
\end{eqnarray}
\end{subequations}
by using the relations
\begin{eqnarray}
&&
r^{\ }_{\delta L}=
\left(
{i}
 v^{\ }_1
-v^{\ }_2
\right)
\delta L,
\quad
t^{\ }_{\delta L}= 
1
+
{i}
\left(
v^{\ }_0
+
E
\right)
\delta L,
\nonumber \\
&&
r^{\prime}_{\delta L}=
\left(
{i}
 v^{\ }_1
+v^{\ }_2
\right)
\delta L,
\quad
t^{\prime}_{\delta L}=
1
+
{i}
\left(
v^{\ }_0
+
E
\right)
\delta L,
\hphantom{AAAA}
\end{eqnarray}
which are valid up to first order in the disorder potentials
$v^{\ }_{0,1,2}$,
and by expanding the right-hand side of 
Eq.\ (\ref{eq:composition law reflection amplitude})
to the same order. We note in passing that
the mean free path (\ref{eq: def ell}) is simply given by
\begin{eqnarray}
\ell=
\frac{1}{2\left(g^{\ }_1+g^{\ }_2\right)}
\label{eq: ell to second order in Born approximation}
\end{eqnarray}
in the Born approximation.
For later convenience, we have continued the energy $E$ to the 
upper part of the complex
plane,
$
E=\mathcal{E}^{\prime}
+{i}\mathcal{E}^{\prime\prime},
$
$\mathcal{E}^{\prime\prime}\geq0$.
The continuous Langevin process 
(\ref{eq: Langevin eq for R and phi})
can also be formulated as the
Fokker-Planck equation obeyed by the joint probability distribution
function $P(R,\phi;L)$.
Using the standard methods of chapter 3 in 
Ref.\ \onlinecite{Zinnjustin89}
say,
one finds\cite{footnote dicrete Langevin process}
\begin{subequations}
\label{eq: master equation in terms of R and phi}
\begin{eqnarray}
&&
\frac{\partial P}{\partial t}
=F^{\ }_{0}P
-F^{\ }_{R}\frac{\partial P}{\partial R}
-F^{\ }_{\phi}\frac{\partial P}{\partial \phi}
\\
&&
\hphantom{\frac{\partial P}{\partial t}=}
+\frac{1}{2}G^{\ }_{RR}\frac{\partial^{2}P}{\partial R^{2}}
+G^{\ }_{R\phi}\frac{\partial^{2}P}{\partial R \partial\phi }
+\frac{1}{2}G^{\ }_{\phi\phi}\frac{\partial^{2}P}{\partial \phi^{2}},
\hphantom{AAA}
\nonumber
\end{eqnarray}
where
\begin{eqnarray}
F^{\ }_{0}(R,\phi)&=&
\hphantom{+}2\omega 
+2(2R-1)
-6 \zeta R \cos 2\phi,
\nonumber \\
F^{\ }_{R}(R,\phi)&=&
-2\omega R
-(R-1)(5R-1)
\nonumber \\
&&
+6 \zeta R(R-1) \cos 2\phi ,
\\
F^{\ }_{\phi}(R,\phi)&=&
+\varepsilon
+\frac{\zeta}{2} \left(5R+4+\frac{1}{R}\right)\sin 2\phi,
\nonumber
\end{eqnarray}
and
\begin{eqnarray}
G^{\ }_{RR}(R,\phi)&=&
2R(1-R)^{2}
-2\zeta R(1-R)^{2}
\cos 2\phi,
\nonumber \\
G^{\ }_{R\phi}(R,\phi)&=&
\zeta  (1-R^{2})\sin 2\phi,
\\
G^{\ }_{\phi\phi}(R,\phi)&=&
4\zeta^{\ }_{0}
+
\frac{1}{2}
\left( R+2+\frac{1}{R}\right)
\nonumber \\
&&
+
\frac{\zeta}{2}
\left( R+2+\frac{1}{R}\right)
\cos 2\phi.
\nonumber
\end{eqnarray} 
\end{subequations}
We have introduced the short-hand notations
\begin{eqnarray}
\begin{array}{lll}
g^{\ }_{\pm}:=g^{\ }_{1}\pm g^{\ }_{2},
&
\qquad
&
t:=L/\ell,
\\
\varepsilon:=\mathcal{E}^{\prime}/g^{\ }_{+},
&
\qquad
&
\omega:= \mathcal{E}^{\prime\prime}/g^{\ }_{+},
\\
\zeta:=g^{\ }_{-}/g^{\ }_{+},
&
\hphantom{AAAAAA}
&
\zeta^{\ }_{0}:=g^{\ }_{0}/g^{\ }_{+}.
\end{array}
\end{eqnarray}
Observe that the dependence on the polar angle $\phi$ arises solely from 
$\cos 2\phi$ 
and 
$\sin 2\phi$
in the Fokker-Planck equation 
(\ref{eq: master equation in terms of R and phi}).
Thus, provided initial and boundary conditions are also periodic 
on the interval $[0,\pi[$,
the joint probability distribution $P(R,\phi;L)$
is periodic on the interval $[0,\pi[$ 
even though the phase $\phi$ is originally defined
modulo $2\pi$. The periodicity $\pi$ is a consequence of
the Langevin process (\ref{eq: Langevin eq for R and phi})
being invariant under
$\sqrt{R}\to-\sqrt{R}$
and 
$\phi\to\phi+\pi$.
A dimerization gap $\left\langle v^{\ }_2(y)\right\rangle=\Delta$
produces the changes
\begin{subequations}
\begin{eqnarray}
&&
F^{\ }_0(R,\phi)\to
F^{\ }_0(R,\phi)
+
\frac{\Delta}{g^{\ }_+}
(R-1)\frac{1}{\sqrt{R}}\cos\phi,
\\
&&
F^{\ }_R(R,\phi)\to
F^{\ }_R(R,\phi)
+
\frac{\Delta}{g^{\ }_+}
(R-1)\sqrt{R}\cos\phi,
\\
&&
F^{\ }_\phi(R,\phi)\to
F^{\ }_\phi(R,\phi)
+
\frac{\Delta}{2g^{\ }_+}
(R+1)\frac{1}{\sqrt{R}}\sin\phi.
\hphantom{AAAA}
\end{eqnarray}
\end{subequations}

The polar decomposition (\ref{eq: polar decomposition S+M})
suggests the change of variable
\begin{equation}
\sqrt{R}=\tanh |x|,
\label{eq: R in terms x}
\end{equation}
in terms of which the Langevin process
(\ref{eq: Langevin eq for R and phi})
becomes 
\begin{subequations}
\label{eq: Langevin eq for x and phi}
\begin{eqnarray}
\frac{{d}x}{{d}L}&=&
-\mathcal{E}^{\prime\prime} \sinh 2 x
+v_{1}\sin\phi-v_{2}\cos\phi ,
\\
\frac{{d}\phi}{{d}L}&=&
2
\left(
\mathcal{E}^{\prime}
+v_{0}
\right)
+\frac{2}{\tanh 2x} 
\left( 
 v_{1}\cos\phi
+v_{2}\sin\phi
\right).
\hphantom{AA}
\nonumber\\
&&
\end{eqnarray}
\end{subequations}
In turn, the joint probability distribution
function $W(x,\phi;L)$ 
obeys the Fokker-Planck equation
\begin{subequations}
\label{eq: master equation in terms of x and phi}
\begin{eqnarray}
&&
\frac{\partial W}{\partial t}=
 J^{\ }_{0}W
-J^{\ }_{x}\frac{\partial W}{\partial x}
-J^{\ }_{\phi}\frac{\partial W}{\partial \phi}
\\
&&
\hphantom{\frac{\partial W}{\partial t}=}
+\frac{1}{2}K^{\ }_{xx}\frac{\partial^{2}W}{\partial x^{2}}
+K^{\ }_{x\phi}\frac{\partial^{2}W}{\partial x \partial\phi }
+\frac{1}{2}K^{\ }_{\phi\phi}\frac{\partial^{2}W}{\partial \phi^{2}},
\hphantom{AA}
\nonumber
\end{eqnarray}
where
\begin{eqnarray}
J^{\ }_{0}(x,\phi)&=&
\hphantom{+}
2\omega \cosh 2x
+
\frac{1}{\sinh^{2} 2x}
\nonumber \\
&&
-
\zeta 
\left(
\frac{1}{\sinh^{2} 2x}
+
\frac{2}{\tanh^{2} 2x}
\right)\cos 2\phi,
 \\
J^{\ }_{x}(x,\phi)&=&
-\omega \sinh 2x 
+\frac{1}{2}
\frac{1}{\tanh 2x}
-\frac{3}{2}
\frac{\zeta}{\tanh 2x}
 \cos 2\phi,
\nonumber \\
J^{\ }_{\phi}(x,\phi)&=&
\hphantom{+}
\varepsilon
+
\zeta  
\left(
\frac{1}{\sinh^{2} 2x}
+
\frac{3}{\tanh^{2} 2x}
\right)\sin 2 \phi,
\nonumber
\end{eqnarray}
and
\begin{eqnarray}
K^{\ }_{xx}(x,\phi)&=&
\frac{1}{2}
\left(
1
-
\zeta\cos 2\phi 
\right),
\nonumber \\
K^{\ }_{x\phi}(x,\phi)&=&
\frac{\zeta}{\tanh 2x}
\sin 2 \phi,
\\
K^{\ }_{\phi\phi}(x,\phi)&=&
4\zeta^{\ }_{0}
+\frac{2}{\tanh^{2} 2x}
+\zeta
\frac{2}{\tanh^{2} 2x}
\cos 2\phi.
\hphantom{AA}
\nonumber
\end{eqnarray} 
\end{subequations}
The Langevin process (\ref{eq: Langevin eq for x and phi})
or the Fokker-Planck equation 
(\ref{eq: master equation in terms of x and phi})
are invariant under $x\to-x$ and $\phi\to\phi+\pi$.

Equations (\ref{eq: master equation in terms of R and phi})
or        (\ref{eq: master equation in terms of x and phi})
are, from a conceptual point of view, the main result of this paper
as they determine under what conditions one-parameter scaling holds
in a 1D weakly disordered quantum wire.
For example, Eq.\ (\ref{eq: master equation in terms of x and phi})
encodes the functional renormalization group flow of the 
joint probability distribution of the radial coordinate 
$|x|=\mathrm{arctanh}\, \sqrt{R}$ and phase $\phi$
of the reflection coefficient $r$. 
If it is the property that the functional renormalization group flow of the 
probability distribution of $x$ depends solely on
the dimensionless ratio $L/\ell$ aside from $x$,
which is understood as one-parameter scaling for $x$,\cite{Shapiro87}
then two examples of one-parameter scaling can be constructed from
Eq.\ (\ref{eq: master equation in terms of x and phi})
as follows.
First, the Fokker-Planck (DMPK) equation (\ref{eq: FP std limit})
follows from insertion of the ansatz
\begin{eqnarray}
W(x,\phi;L)=
\frac{1}{2\pi}
\mathcal{X}(x;L)
\end{eqnarray}
into Eq.\ (\ref{eq: master equation in terms of x and phi})
with $\omega=0$ 
and from integration over $\phi\in[0,2\pi[$ of the resulting equation.
Second, the Fokker-Planck (DMPK) equation (\ref{eq: FP chiral limit})
follows from insertion of the ansatz
\begin{eqnarray}
W(x,\phi;L)=
\frac{1}{2}\left[\delta(\phi-0)+\delta(\phi-\pi)\right]
\mathcal{X}(x;L)
\hphantom{AA}
\end{eqnarray}
with $g^{\ }_0=g^{\ }_1=\omega=\varepsilon=0$
into Eq.\ (\ref{eq: master equation in terms of x and phi})
and from integration over $\phi$ of the resulting equation.
One-parameter scaling is not the rule
for generic initial and boundary conditions of
Eq.\ (\ref{eq: master equation in terms of x and phi})
and for generic values of $\varepsilon$ and $g^{\ }_{0,1,2}$
due to the absence of a diffusive regime in 1D.\cite{Brouwer03}
This is most clearly seen by the fact that the localization
length $\xi$, as defined by the typical dependence of the conductance 
$g=\mathrm{sech}^2\,x$ on $L\gg\ell$, is finite,
$\xi=2\ell$,
for the standard symmetry class,
but diverges in the chiral limit
$g^{\ }_0=g^{\ }_1=\omega=\varepsilon=0$.
Hence, the localization length must depend on a second microscopic parameter
aside from $\ell$ for generic values of 
$\varepsilon$ and $g^{\ }_{0,1,2}$.
Equations (\ref{eq: master equation in terms of R and phi})
and 
(\ref{eq: master equation in terms of x and phi})
describe the full crossover from the chiral to the standard class
with no restrictions on the values of $x$ and $L$.

\begin{figure}[t]
\begin{center}
\begin{picture}(180,170)(0,10) 
\thicklines
\put( +60,+70 ){\vector(0,1){100}}
\put( 0,+55 ){\vector(4,1){180}}
\put( 30,+90 ){\vector(3,-2){120}}

\thinlines
\put( 20,60 ){\line(0,1){40}}
\put( 100,80 ){\line(0,1){40}}

\put( 80,20 ){\line(0,1){70}}
\put( 120,30 ){\line(0,1){70}}
\put( 160,40 ){\line(0,1){70}}

\put( 60,15 ){\line(4,1){120}}

\put( 20,100 ){\line(4,1){80}}

\put( 80,90 ){\line(4,1){80}}

\put( 95,10 ){\line(-3,2){90}}
\put( 85,90 ){\line(3,-2){90}}
\put( 30,160 ){\line(3,-2){90}}

\put(80,80){\line(1,1){13}}
\put(80,75){\line(1,1){20}}
\put(80,70){\line(1,1){26}}
\put(80,65){\line(1,1){32}}
\put(80,60){\line(1,1){40}}
\put(80,55){\line(1,1){46}}
\put(80,50){\line(1,1){53}}
\put(80,45){\line(1,1){59.5}}
\put(80,40){\line(1,1){66.5}}
\put(80,35){\line(1,1){73}}
\put(80,30){\line(1,1){80}}
\put(80,25){\line(1,1){80}}
\put(80,20){\line(1,1){80}}

\put(160,95){\line(-1,-1){73.5}}
\put(160,90){\line(-1,-1){66.5}}
\put(160,85){\line(-1,-1){59.5}}
\put(160,80){\line(-1,-1){53}}
\put(160,75){\line(-1,-1){46}}
\put(160,70){\line(-1,-1){40}}
\put(160,65){\line(-1,-1){33}}
\put(160,60){\line(-1,-1){26.5}}
\put(160,55){\line(-1,-1){19.5}}
\put(160,50){\line(-1,-1){13}}

\put(30,95){\line(1,-1){15}}
\put(30,100){\line(1,-1){30}}
\put(30,105){\line(1,-1){45}}
\put(30,110){\line(1,-1){60}}
\put(30,115){\line(1,-1){75}}
\put(30,120){\line(1,-1){90}}
\put(30,125){\line(1,-1){90}}
\put(30,130){\line(1,-1){90}}
\put(30,135){\line(1,-1){90}}
\put(30,140){\line(1,-1){90}}
\put(30,145){\line(1,-1){90}}
\put(30,150){\line(1,-1){90}}
\put(30,155){\line(1,-1){90}}
\put(30,160){\line(1,-1){90}}

\put(120,95){\line(-1,1){15}}
\put(120,90){\line(-1,1){30}}
\put(120,85){\line(-1,1){45}}
\put(120,80){\line(-1,1){60}}
\put(120,75){\line(-1,1){75}}

\put(170,105){$\zeta$}
\put(48,165){$\zeta^{\ }_{0}$}
\put(150,15){$\varepsilon$}
\put(65,11){$+\infty$}
\put(13,50){$-1$}
\put(58,60){$0$}
\put(93,70){$+1$}
\put(62,113){$1$}
\put(21,65){Ch1}
\put(102,85){Ch2}
\put(21,105){KW1}
\put(102,125){KW2}
\put(20,60){\circle{4}}
\put(100,80){\circle{4}}
\put(20,100){\circle*{4}}
\put(100,120){\circle*{4}}

\end{picture}
\caption{
Parameter space for the Fokker-Planck equations
(\ref{eq: master equation in terms of R and phi})
or
(\ref{eq: master equation in terms of x and phi}).
The point Ch1 has the coordinates
$\varepsilon=\zeta^{\ }_0=0$, $\zeta=-1$.
It represents the chiral symmetry class.
The point KW1 has the coordinates
 $\varepsilon=0$, $\zeta^{\ }_0=-\zeta= 1$.
It represents the Kappus-Wegner anomaly.
The two shaded planes represent 
the standard symmetry class.
The regions $\zeta > 0$ and $\zeta < 0$
are equivalent in that they are related by 
a rotation by an angle $\pi$
around $\sigma^{\ }_3$ of
the Pauli matrices in
the Hamiltonian (\ref{eq: def our Anderson localization problem}).
Especially, this rotation
relates Ch1 and KW1 to the points Ch2 and KW2, respectively.
\label{fig: parms}
}
\end{center}
\end{figure}
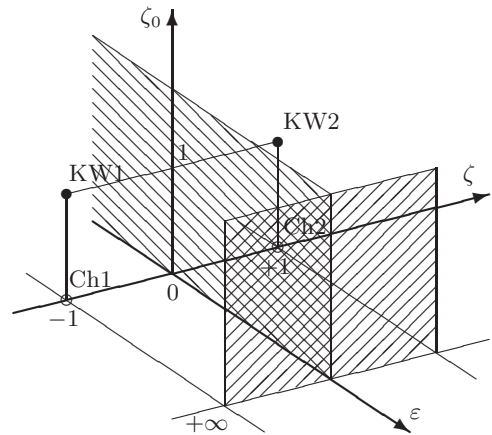

For any given absorption $\omega\geq0$, the parameter space 
$(\zeta,\zeta^{\ }_0,\varepsilon)$ 
with $\zeta^{\ }_0 \geq0$ and $\zeta,\varepsilon\in\mathbb{R}$
of
Eqs.\ (\ref{eq: master equation in terms of R and phi})
and   (\ref{eq: master equation in terms of x and phi})
is three-dimensional and is depicted in Fig.\ \ref{fig: parms}.
Remarkably, the parameter $\zeta$ always enters
Eqs.\ (\ref{eq: master equation in terms of R and phi})
and   (\ref{eq: master equation in terms of x and phi})
as a prefactor to $\cos2\phi$ or $\sin2\phi$ and conversely
$\cos2\phi$ or $\sin2\phi$ are always multiplied by $\zeta$.
As we shall see in Sec.\ 
\ref{sec: Solutions to the functional renormalization group equations}
this property turns out to be crucial in our study of the
crossover between the chiral and standard symmetry classes.

Equation (\ref{eq: master equation in terms of R and phi})
was derived in the standard class by 
Abrikosov, Melnikov, and Kumar, and by Rammal and Doucot 
among others
starting from the continuous nonrelativistic Schr\"odinger 
equation.\cite{Abrikosov81,Melnikov81,Kumar85,Rammal87}
The large $x$ limit
\begin{eqnarray}
\frac{\partial W}{\partial t}
&\approx&
\hphantom{+}\left(
\omega e^{2x}
-
2\zeta
\cos 2\phi 
\right)
W
\nonumber\\
&&
+
\left(
\frac{\omega}{2}e^{2x}
-\frac{1}{2}
+\frac{3}{2}\zeta
 \cos 2\phi
\right)
\frac{\partial W}{\partial x}
\nonumber\\
&&
+
\left(
-
\varepsilon
-
3\zeta \sin 2 \phi  
\right)
\frac{\partial W}{\partial \phi}
\nonumber\\
&&
+
\frac{1}{4}
\left(
1
-
\zeta\cos 2\phi 
\right)
\frac{\partial^2W}{\partial x^2}
\nonumber\\
&&
+\zeta
\sin (2 \phi)  
\frac{\partial^2W}{\partial x\partial\phi}
\nonumber\\
&&
+
\left(
2\zeta_{0}
+1
+\zeta
\cos 2\phi 
\right)
\frac{\partial^2W}{\partial \phi^2}
\label{eq: large x limit FP x phi all derivatives to right}
\end{eqnarray}
of Eq.\ (\ref{eq: master equation in terms of x and phi})
was derived by Schomerus and Titov when 
$\omega=0$ 
and 
$g^{\ }_0=g^{\ }_1$
starting from a Langevin process for the eigenfunctions
of the discrete Schr\"odinger equation on a chain. 
\cite{Schomerus03a,Schomerus03b}

\section{
Solutions to the functional renormalization group equations
        }
\label{sec: Solutions to the functional renormalization group equations}

Exact and approximate solutions to the 
Fokker-Planck equation (\ref{eq: master equation in terms of R and phi})
are derived in this section. 
We will consider three regimes of
parameter in Fig.\ \ref{fig: parms}.
The first regime corresponds to the chiral symmetry class.
The second regime corresponds to the standard symmetry class.
The third regime describes a specific path connecting the two
symmetry classes. 

The chiral symmetry class is given by the condition
\begin{subequations}
\label{eq: def chiral regime parameters}
\begin{eqnarray}
\varepsilon=0,
\label{eq: def chiral regime parameters a}
\end{eqnarray}
which defines the band center of the energy eigenvalue spectrum 
of Hamiltonian (\ref{eq: def our Anderson localization problem a})
when
\begin{eqnarray}
\zeta=-1,
\qquad
\zeta^{\ }_0=0,
\qquad
(g^{\ }_0=g^{\ }_1=0),
\label{eq: def chiral regime parameters b}
\end{eqnarray}
\end{subequations}
i.e., when Hamiltonian (\ref{eq: def our Anderson localization problem a})
anticommutes with $\sigma^{\ }_1$ as in
Eq.\ (\ref{eq: def chiral symmetry for H}).
The standard symmetry class occurs whenever
\begin{subequations}
\label{eq: strong+weak def of std limit}
\begin{eqnarray}
\zeta=0
\qquad
(g^{\ }_1=g^{\ }_2)
\label{eq: strong def of std limit}
\end{eqnarray}
or
\begin{eqnarray}
1/\zeta^{\ }_0=0,
\label{eq: yet another def of std limit}
\end{eqnarray}
or
\begin{eqnarray}
1/|\varepsilon|=0
\label{eq: weak def of std limit}
\end{eqnarray}
\end{subequations}
since any one of these three conditions decouples $R$ from $\phi$
in Eq.\ (\ref{eq: master equation in terms of R and phi})
and is consistent with a stationary probability distribution of $\phi$
which is uniformly distributed. 
In this paper we shall only consider the crossover between the
chiral and standard symmetry classes along the line
\begin{eqnarray}
\zeta=-1, 
\qquad
\zeta_{0}=0,
\qquad
0<\varepsilon<\infty.
\label{eq: crossover regime weak if a}
\end{eqnarray}


We must naturally choose the initial and
boundary conditions to be imposed on the solutions to the
Fokker-Planck equation (\ref{eq: master equation in terms of R and phi}).
Our choice is motivated by the computation of the probability distribution
of the LDOS that we will carry out in Sec.\
\ref{sec: Probability distribution of the LDOS}.
Two geometries depicted in Fig.\ \ref{fig: BC for LDOS}
will be considered in Sec.\
\ref{sec: Probability distribution of the LDOS}.
The relevant geometries for the probability distribution 
of the reflection coefficient $r$ that enter the computation of the
LDOS
are depicted in
Fig.\ \ref{fig: geometries}.
In Figs.\ \ref{fig: geometries}a and \ref{fig: geometries}b
the disordered quantum wire is closed on the right-hand side and
connected to a reservoir on the left-hand side.
In Fig.\ \ref{fig: geometries}a conservation of probability holds.
In Fig.\ \ref{fig: geometries}b conservation of probability does not hold
due to the presence of an absorption encoded by a finite imaginary part
$\mathcal{E}^{\prime\prime}=\omega g^{\ }_+$
of $E$.
In Fig.\ \ref{fig: geometries}c
the disordered quantum wire is open at both ends and flux conservation
(no absorption) is assumed.
We choose the ideal initial condition $r=1$ and $r=0$ when $L=0$
for the semiopen and fully open geometries of Fig.\ \ref{fig: geometries},
respectively.
In the chiral symmetry class, we demand that the phase of $r$
is either 0 or $\pi$ with probability 1/2
initially. In the standard symmetry class, we demand
that the phase of $r$ is uniformly distributed in the interval $[0,2\pi[$
initially. In view of the periodicity of the
Fokker-Planck equation (\ref{eq: master equation in terms of R and phi})
we always impose initial conditions with a periodicity of $\pi$.

We shall now construct exact solutions to the
Fokker-Planck equation (\ref{eq: master equation in terms of R and phi})
in the chiral and standard symmetry classes, respectively,
for the three geometries of Fig.\ \ref{fig: geometries}.
We shall also solve the
Fokker-Planck equation (\ref{eq: master equation in terms of R and phi})
in the crossover regime (\ref{eq: crossover regime weak if a})
for the geometry of Fig. \ref{fig: geometries}a.
We could not construct exact solutions to the
Fokker-Planck equation (\ref{eq: master equation in terms of R and phi})
in the crossover between the chiral and standard symmetry classes 
for the geometries
of Figs.\ \ref{fig: geometries}b
and       \ref{fig: geometries}c.
We shall construct, however, an approximate solution to the
Fokker-Planck equation (\ref{eq: master equation in terms of R and phi})
in a crossover regime
that interpolates between the exact solutions for the geometries
of Figs.\ \ref{fig: geometries}b
and       \ref{fig: geometries}c.
The accuracy of this approximate solution will be discussed in 
Sec.\ \ref{sec: Validity of the separation ansatz for the crossover regime}.

\subsection{
Disordered quantum wire closed on the right-hand side without absorption
           }
\label{subsec: disordered wire closed on the right-hand side without absorption}

In geometry \ref{fig: geometries}a,
an incoming plane wave from an ideal lead 
is perfectly reflected by the disordered quantum wire. Hence, the
reflection coefficient $r$ of the
disordered quantum wire must be a pure phase,
$rr^*=1$, for all $t=L/\ell$.
We thus insert the separation-of-variable ansatz
\begin{subequations}
\begin{eqnarray}
P(R,\phi;t)=
\delta(R-1)\,\Phi(\phi;t)
\end{eqnarray}
into the
Fokker-Planck equation (\ref{eq: master equation in terms of R and phi})
that, after integration over $R$, reduces to
\begin{eqnarray}
\frac{\partial\Phi}{\partial t}
&=&
-2\zeta \cos(2\phi) \Phi
-
\left[
\varepsilon
+3\zeta \sin(2\phi)
\right]
\frac{\partial\Phi}{\partial\phi}
\nonumber \\
&&
+
\left[
 2\zeta^{\ }_{0}
+1
+\zeta \cos(2\phi)
\right]
\frac{\partial^{2}\Phi}{\partial\phi^{2}}.
\label{eq: FP if closed at one end and no absorption}
\end{eqnarray}
\end{subequations}
By assumption, we have switched off the absorption $\omega=0$.
Our choice of initial condition shall depend on the proximity
to the parameter regimes
(\ref{eq: def chiral regime parameters})
and
(\ref{eq: strong+weak def of std limit})
but shall always be periodic on the interval $[0,\pi[$.

\subsubsection{
Chiral symmetry class
              }
\label{subsubsec: Chiral symmetry class a}

In the chiral symmetry class
(\ref{eq: def chiral regime parameters})
the Fokker-Planck equation
(\ref{eq: FP if closed at one end and no absorption})
reduces to
\begin{subequations}
\begin{eqnarray}
\frac{\partial\Phi}{\partial t}
&=&
2 \cos(2\phi)\Phi
+3 \sin(2\phi)
\frac{\partial\Phi}{\partial\phi}
+
\left[
1-\cos(2\phi)
\right]\frac{\partial^{2}\Phi}{\partial\phi^{2}}
\nonumber \\
\label{eq: FP if closed at one end and no absorption ch}
\end{eqnarray}
and has the normalized, stationary, and periodic solution 
\begin{eqnarray}
\Phi(\phi)=
\frac{1}{2}
\delta(\phi-0)
+
\frac{1}{2}
\delta(\phi-\pi)
\label{eq: sum of delta functions}
\end{eqnarray}
\end{subequations}
whenever the phase $\phi$ is $0$ or $\pi$ with probability $1/2$ initially.
The same conclusion could have been anticipated by inspection of
the Langevin process (\ref{eq: Langevin eq for R and phi})
or from the fact that the chiral symmetry class
(\ref{eq: def chiral regime parameters})
imposes the condition
that the scattering matrix is Hermitian, 
see Eq.\ (\ref{eq: chiral symmetry S}).

\subsubsection{
Standard symmetry class
              }
\label{subsubsec: Standard symmetry class a}

In the standard symmetry class 
(\ref{eq: strong def of std limit}),
$\cos 2\phi$ and $\sin 2\phi$ drop out from 
the Fokker-Planck equation 
(\ref{eq: FP if closed at one end and no absorption}):
\begin{subequations}
\begin{eqnarray}
\frac{\partial \Phi}{\partial t}=
-\varepsilon\frac{\partial\Phi}{\partial \phi}
+(2\zeta^{\ }_0+1)\frac{\partial^{2}\Phi}{\partial\phi^{2}}.
\label{eq: FP if closed at one end and no absorption std}
\end{eqnarray}
Its normalized, stationary, and periodic solution is
\begin{eqnarray}
\Phi(\phi)=\frac{1}{2\pi}
\label{eq: solution std limit semi-open geometry}
\end{eqnarray}
\end{subequations}
whenever the phase $\phi$ of $r$
is uniformly distributed initially.
More generally, 
the solution to
Eq.\ (\ref{eq: FP if closed at one end and no absorption std})
converges to the uniform probability distribution
(\ref{eq: solution std limit semi-open geometry}) 
as soon as  $t\gg1$ for any given initial probability distribution.
This is also true for the standard symmetry class
(\ref{eq: yet another def of std limit})
and
(\ref{eq: weak def of std limit})
as $\cos(2\phi)$ and $\sin(2\phi)$ are ineffective in these limits.

When $g^{\ }_{1}\neq g^{\ }_{2}$ and $|\varepsilon|\lesssim1$
the stationary probability distribution of $r$ need not be uniform in 
$[0,2\pi[$.
For example, consider the case
$\zeta=+1$,
$\varepsilon =0$,
$\zeta^{\ }_{0}=\frac{1}{2}(1+\zeta)=1$,
for which the Fokker-Planck equation 
(\ref{eq: FP if closed at one end and no absorption})
reduces to
\begin{subequations}
\label{eq: band center anomaly}
\begin{eqnarray}
\frac{\partial \Phi}{\partial t}
&=&
-2 \cos(2\phi)\Phi
-3 \sin(2\phi)
\frac{\partial \Phi}{\partial\phi}
\nonumber \\
&&
+
\left[
3+\cos(2\phi)
\right]
\frac{\partial^{2}\Phi}{\partial\phi^{2}}
\label{eq: band center anomaly a}
\end{eqnarray}
and has the stationary and normalized solution\cite{Schomerus03a,Kappus81}
\begin{eqnarray}
\Phi(\phi)&=&
\frac{\sqrt{2\pi}}{\Gamma^{2}(1/4)\sqrt{1+\cos^{2}\phi}}
\label{eq: band center anomaly b}
\end{eqnarray}
with a periodicity of $\pi$.
Here $\Gamma(z)$ denotes the value of the gamma function for $z\in\mathbb{C}$.
Equation (\ref{eq: band center anomaly b})
is a special case of the stationary solution
\begin{eqnarray}
\Phi(\phi)
&\propto&
\frac{1}{\sqrt{2\zeta_{0}+1+\zeta \cos(2\phi)}}
\label{eq: Titov sol}
\end{eqnarray}
\end{subequations}
to
Eq.\ (\ref{eq: FP if closed at one end and no absorption})
with $\varepsilon=0$
as shown by Titov.\cite{Titov04}
In the chiral limit $\zeta \to -1$ and $\zeta^{\ }_{0}\to 0$,
this solution becomes unnormalizable and we have to use
Eq.\ (\ref{eq: sum of delta functions})
as a stationary solution instead.

\subsubsection{
Crossover regime
              }
\label{subsubsec: Crossover regime a}

For simplicity, we consider the crossover regime
(\ref{eq: crossover regime weak if a}).\cite{footnote-cross}
The chiral symmetry class 
(\ref{eq: def chiral regime parameters})
is reached when $\varepsilon=0$.
The standard symmetry class 
(\ref{eq: weak def of std limit})
is reached when 
$|\varepsilon|\gg1$.
In the crossover regime,
the Fokker-Planck equation 
(\ref{eq: FP if closed at one end and no absorption})
reduces to
\begin{eqnarray}
\frac{\partial \Phi}{\partial t}
&=&
-\varepsilon \frac{\partial\Phi}{\partial \phi}
+2\frac{\partial}{\partial \phi}
\left(\sin\phi \frac{\partial}{\partial\phi}\sin \phi 
\right)\Phi.
\label{eq: chiral 1 FP with no absorption and R=1}
\end{eqnarray}
A stationary solution to 
Eq.\ (\ref{eq: chiral 1 FP with no absorption and R=1})
is constructed from the solution 
\begin{eqnarray}
\Phi(\phi)=
\frac{\mathcal{N}}{\sin\phi}
\int_{\phi}^{\pi} {d}\phi^{\prime}\,
\frac{
e^{+(\varepsilon/2)(\cot \phi^{\prime}-\cot\phi)}
}
{\sin \phi^{\prime}},
\label{eq: sol cross if geo c}
\end{eqnarray}
valid for $\varepsilon > 0$ and
on the interval $[0,\pi[$, 
by periodic extension to the interval $[0,2\pi[$.
The constant $\mathcal{N}$ is chosen so that the solution
(\ref{eq: sol cross if geo c}) is properly normalized.
The solution
interpolates between
Eqs.\ (\ref{eq: sum of delta functions})
and
      (\ref{eq: solution std limit semi-open geometry})
as $\varepsilon\to0$ and $\varepsilon\to\infty$, respectively,
as it should be.
Equation (\ref{eq: sol cross if geo c})
is an important intermediary result
as we will use it to construct an approximation
to the probability distributions for the conductance and LDOS
in the crossover regime that interpolate between the exact chiral
and standard limiting probability distributions.

\subsection{
Disordered quantum wire closed on the right-hand side with absorption
           }
\label{subsec: Disordered quantum wire closed on the right-hand side 
               with absorption}

In geometry \ref{fig: geometries}b,
an incoming plane wave from an ideal lead
is not perfectly reflected by the disordered quantum wire due to a finite
absorption $\omega= \mathcal{E}^{\prime\prime}/g^{\ }_+$.
Hence, the reflection coefficient $r$ of the
disordered quantum wire is not a pure phase anymore,
$r=\sqrt{R}\exp({i}\phi)$.

\subsubsection{
Chiral symmetry class
              }
\label{subsubsec: Chiral symmetry class b}

In the chiral symmetry class (\ref{eq: def chiral regime parameters})
the phase $\phi$ and squared magnitude $R$ of the reflection coefficient $r$
must necessarily separate since $0\leq r\leq1$ is necessarily real valued,
\begin{eqnarray}
P(R,\phi;t)=
\left[
\frac{1}{2}
\delta(\phi-  0)
+
\frac{1}{2}
\delta(\phi-\pi)
\right]
\mathcal{R}(R;t).
\hphantom{AA}
\label{eq: separation variable chiral geometry b}
\end{eqnarray}
After insertion of
Eq.\ (\ref{eq: separation variable chiral geometry b})
into the Fokker-Planck equation
(\ref{eq: master equation in terms of R and phi})
one finds that $\mathcal{R}(R;t)$ obeys
\begin{subequations}
\begin{eqnarray}
\frac{\partial \mathcal{R}}{\partial t}
&=&
\hphantom{+}\Big[2\omega +2(3R-2)\Big]\mathcal{R}
\nonumber \\
&&
+\Big[2\omega R + 3 (1-R)(1-3R) \Big]
\frac{\partial \mathcal{R}}{\partial R}
\nonumber \\
&&
+2R(1-R)^{2}\frac{\partial^{2} \mathcal{R}}{\partial R^{2}}
\label{eq: FP chiral geometry b}
\end{eqnarray}
with the stationary solution
\begin{eqnarray}
\mathcal{R}(R)=
\mathcal{N}
\frac{\exp\left(-\frac{\omega }{1-R}\right)}
{\sqrt{R}(1-R)}.
\label{eq: solution R for closed wire with absorption (chiral)}
\end{eqnarray}
\end{subequations}
The constant $\mathcal{N}$ is chosen so that the solution
(\ref{eq: solution R for closed wire with absorption (chiral)})
is properly normalized (this is always possible for $\omega >0$).
The stationary solution 
(\ref{eq: solution R for closed wire with absorption (chiral)})
has a square-root singularity at $R=0$,
an essential singularity at $R=1$, and
is normalizable for $R\in [0,1]$.
We shall see below that the probability for $R$ to be in the vicinity of
0 is enhanced in the chiral symmetry class 
(\ref{eq: def chiral regime parameters})
compared to the standard symmetry class
(\ref{eq: strong+weak def of std limit}).

\subsubsection{
Standard symmetry class
              }
\label{subsubsec: Standard symmetry class b}

In the standard symmetry class (\ref{eq: strong+weak def of std limit})
the phase $\phi$ and squared magnitude $R$ of the reflection coefficient $r$
can reasonably be taken to separate in view of the initial condition of
a uniformly distributed phase $\phi$,
\begin{eqnarray}
P(R,\phi;t)=
\frac{1}{2\pi}
\mathcal{R}(R;t).
\label{eq: separation variable std geometry b}
\end{eqnarray}
After insertion of the separation-of-variable ansatz
(\ref{eq: separation variable std geometry b})
into the Fokker-Planck equation
(\ref{eq: master equation in terms of R and phi})
one finds that $\mathcal{R}(R;t)$ obeys
\begin{subequations}
\begin{eqnarray}
\frac{\partial \mathcal{R}}{\partial t}&=& 
\hphantom{+}\Big[ 2\omega +2(2R-1) \Big]\mathcal{R}
\nonumber \\
&&
+
\Big[ 
2\omega R
+
(1-R)(1-5R)
 \Big]
 \frac{\partial \mathcal{R}}{\partial R}
\nonumber\\
&&
+
R(1-R)^{2}\frac{\partial^{2} \mathcal{R}}{\partial R^{2}}
\label{eq: FP standard limit closed both sides with abs}
\end{eqnarray}
with the stationary and normalized solution
\begin{eqnarray}
\mathcal{R}(R)=
2\omega e^{2\omega}
\frac{\exp\left(-\frac{2\omega }{1-R}\right)}{(1-R)^{2}}.
\label{eq: solution R for closed wire with absorption (std)}
\end{eqnarray}
\end{subequations}
The stationary solution
(\ref{eq: solution R for closed wire with absorption (std)})
has an essential singularity when $R=1$.
The stationary solution
(\ref{eq: solution R for closed wire with absorption (std)})
was found by Pradhan and Kumar in 
Ref.\ \onlinecite{Pradhan94}.
When the initial condition on the probability distribution of $\phi$ 
is not that of a uniform distribution,
it is presumed that the deviations of the solution to the
Fokker-Planck equation
(\ref{eq: master equation in terms of R and phi})
from the separation-of-variable ansatz 
(\ref{eq: separation variable std geometry b}) 
with the stationary solution
(\ref{eq: solution R for closed wire with absorption (std)})
are short transients. We do not know of an analytical verification 
of this assumption.

\subsubsection{
Crossover regime
              }
\label{subsubsec: Crossover regime b} 

There is no compelling reason to believe that the squared amplitude
$R$ and the phase $\phi$ of the reflection coefficient $r$ separate in
the crossover regime. However, short of an explicit exact solution
to the Fokker-Planck equation
(\ref{eq: master equation in terms of R and phi})
in the crossover regime,
we shall nevertheless pursue a strategy relying on an approximation
built on a separation-of-variable ansatz
that allows us to interpolate between the chiral and standard symmetry classes.
One possible alternative to the separation-of-variable ansatz that we
tried is a perturbative expansion about the standard symmetry class
(\ref{eq: strong+weak def of std limit}) 
using a Fourier expansion of the joint probability distribution
for $R$ and $\phi$. However, this approach has the drawback that it 
cannot be expected to interpolate all the way to the chiral symmetry class
(\ref{eq: def chiral regime parameters})
in view of the nature of the singularities
characterizing the limiting probability distributions.
For this reason we will not present the perturbative approach in this paper.

Our starting point is the separation-of-variable ansatz
\begin{subequations}
\label{eq: separation-of-variable ansatz in geometry b}
\begin{eqnarray}
P(R,\phi;t)=\mathcal{R}(R;t)\Phi(\phi;t).
\label{eq: separation-of-variable ansatz a}
\end{eqnarray}
We further assume that the probability distribution function for the phase
$\phi$
shows a quick relaxation to its stationary solution, i.e.,
$\Phi(\phi;t)$ is assumed to be $t$-independent,
\begin{eqnarray}
\Phi(\phi;t)=\Phi(\phi).
\label{eq: separation-of-variable ansatz b}
\end{eqnarray}
\end{subequations}
Insertion of Eq.\ (\ref{eq: separation-of-variable ansatz in geometry b})
into the Fokker-Planck equation
(\ref{eq: master equation in terms of R and phi})
followed by an integration over $\phi$ yields
\begin{subequations}
\label{eq: separation ansatz FP for R geometry b}
\begin{eqnarray}
\frac{\partial \mathcal{R}}{\partial t}
&=&
\hphantom{+}\Big[2\omega +2(2R-1)-2\,\zeta \alpha (R-1)  \Big]
\mathcal{R}
\nonumber\\
&&+
\Big[
 2\omega  R
+(1-R)(1-5R)
\nonumber \\
&&
\hphantom{AAAAaaa}
-2\, \zeta \alpha (1-R)(1-2R)
\Big]
\frac{\partial\mathcal{R}}{\partial R}
\nonumber\\
&&+
(1-\zeta \alpha )
R(1-R)^{2}
\frac{\partial^{2}\mathcal{R}}{\partial R^{2}}
\label{eq: resulting ODE for R if separation ansatz}
\end{eqnarray}
where the real constant $\alpha$ is the ratio of
two Fourier expansion coefficients,
\begin{eqnarray}
\alpha&:=&
\frac{\int_{0}^{2\pi}{d}\phi\, \cos(2\phi)\Phi(\phi)}
{\int_{0}^{2\pi}{d}\phi\, \Phi(\phi)}.
\label{eq: def alpha}
\end{eqnarray}
A stationary solution to 
Eq.\ (\ref{eq: resulting ODE for R if separation ansatz})
is 
\begin{eqnarray}
\mathcal{R}(R)=
\mathcal{N}
\frac{\exp\left[-\frac{2\omega }{(1-\zeta \alpha)(1-R)}\right]}
{
\left. R\right. ^{-\zeta \alpha/(1-\zeta \alpha)}
\left(1-R\right)^{2/(1-\zeta \alpha)}}.
\label{eq: solution for R if separation ansatz}
\end{eqnarray}
\end{subequations}
This solution reduces to the known results
in the standard 
[Eq.\ (\ref{eq: solution R for closed wire with absorption (std)})]
and 
chiral 
[Eq.\ (\ref{eq: solution R for closed wire with absorption (chiral)})]
symmetry classes.
In the standard symmetry class (\ref{eq: strong+weak def of std limit})
the probability distribution of the phase is uniform
$\Phi(\phi)=1/2\pi$
and hence 
$\zeta \alpha = 0$,
whereas in the chiral symmetry class (\ref{eq: def chiral regime parameters})
it is a sum of the two delta functions
$
\Phi(\phi)=
\frac{1}{2}
\delta(\phi-0)
+
\frac{1}{2}
\delta(\phi-\pi)
$,
$\alpha = 1$,
$\zeta=-1$.
There remains considerable arbitrariness in the choice of
$\Phi$ because many different $\Phi$ share the same $\alpha$.
In practice we will choose the stationary solution
(\ref{eq: sol cross if geo c})
since we then recover the true solution in the
semiopen geometry
of Fig.\ \ref{fig: geometries}a
as $\omega\to0$.
We postpone to Sec.\ 
\ref{sec: Validity of the separation ansatz for the crossover regime}
the discussion of
the accuracy of the approximation implied by the separation-of-variable
ansatz (\ref{eq: separation ansatz FP for R geometry b})
with the choice (\ref{eq: sol cross if geo c}) 
for the stationary probability distribution of $\phi$.

\subsection{
Disordered quantum wire opened at both ends without absorption
           }
\label{sec: disordered wire opened at both ends without absorption}

In geometry \ref{fig: geometries}c,
an incoming plane wave from an ideal lead 
is partially reflected and partially transmitted
by the disordered quantum wire.
Following Abrikosov in Ref.\ \onlinecite{Abrikosov81}
we trade the squared magnitude $R$ of the reflection coefficient $r$
for the resistance (inverse of conductance)
\begin{subequations}
\begin{eqnarray}
\varrho:=\frac{1}{1-R}\in [1,\infty]
\label{eq: def varrho}
\end{eqnarray}
under which
\begin{eqnarray}
Q(\varrho,\phi;t)=
P(R,\phi;t)\, \frac{{d}R}{{d}\varrho}
\end{eqnarray}
\end{subequations}
obeys the Fokker-Planck equation
\begin{subequations}
\label{eq: master equation in terms of varrho and phi}
\begin{eqnarray}
&&
\frac{\partial Q}{\partial t}
=A^{\ }_{0}Q
-A^{\ }_{\varrho}\frac{\partial Q}{\partial \varrho}
-A^{\ }_{\phi}\frac{\partial Q}{\partial \phi}
\\
&&
\hphantom{\frac{\partial Q}{\partial t}=}
+\frac{1}{2}B^{\ }_{\varrho\varrho}\frac{\partial^{2}Q}{\partial \varrho^{2}}
+B^{\ }_{\varrho\phi}\frac{\partial^{2}Q}{\partial\varrho\partial\phi }
+\frac{1}{2}B^{\ }_{\phi\phi}\frac{\partial^{2}Q}{\partial \phi^{2}},
\hphantom{AAA}
\nonumber
\end{eqnarray}
where
\begin{eqnarray}
A^{\ }_{0}(\varrho,\phi)&=&
\hphantom{+}2\omega (2\varrho-1),
\nonumber \\
A^{\ }_{\varrho}(\varrho,\phi)&=&
-2\omega \varrho (\varrho-1)
-(2\varrho-1),
\\
A^{\ }_{\phi}(\varrho,\phi)&=&
\hphantom{+}\varepsilon
+\frac{\zeta}{2} 
\frac{2\varrho^{2}-2\varrho+1}{\varrho(\varrho-1)}
\sin 2\phi,
\nonumber
\end{eqnarray}
and
\begin{eqnarray}
B^{\ }_{\varrho\varrho}(\varrho,\phi)&=&
2\varrho(\varrho-1)
\left(1-\zeta \cos 2\phi
\right),
\nonumber \\
B^{\ }_{\varrho\phi}(\varrho,\phi)&=&
\zeta (2\varrho-1)\sin 2\phi,
\\
B^{\ }_{\phi\phi}(\varrho,\phi)&=&
4\zeta^{\ }_{0}
+
\frac{1}{2}
\frac{(2\varrho-1)^{2}}{\varrho(\varrho-1)}
\nonumber \\
&&
+
\frac{\zeta}{2}
\frac{(2\varrho-1)^{2}}{\varrho(\varrho-1)}
\cos 2\phi.
\nonumber
\end{eqnarray} 
\end{subequations}

\subsubsection{
Chiral symmetry class
              }
\label{subsubsec: Chiral symmetry class c}

In the chiral symmetry class (\ref{eq: def chiral regime parameters})
insertion of the separation-of-variable ansatz
\begin{eqnarray}
Q(\varrho,\phi;t)=
\left[
\frac{1}{2}
\delta(\phi-0)
+
\frac{1}{2}
\delta(\phi-\pi)
\right]
\mathcal{Q}(\varrho;t)
\end{eqnarray}
into the Fokker-Planck equation
(\ref{eq: master equation in terms of varrho and phi})
yields
\begin{subequations}
\label{eq: FP+initial geometry c chiral}
\begin{eqnarray}
\label{eq: FP c chiral}
\frac{\partial \mathcal{Q}}{\partial t}
&=&
 2\mathcal{Q}
+
3(2\varrho-1)\frac{\partial \mathcal{Q}}{\partial \varrho}
+
2\varrho (\varrho-1)
\frac{\partial^{2} \mathcal{Q}}{\partial \varrho^{2}}
\nonumber\\
&&
\end{eqnarray}
with the initial condition
\begin{eqnarray}
\label{eq: initial cond geometry c chiral}
\mathcal{Q}(\varrho;t=0)=\delta(\varrho-1).
\end{eqnarray}
\end{subequations}
The normalized solution to
Eq.\ (\ref{eq: FP+initial geometry c chiral})
has the integral representation
\begin{subequations}
\label{eq: sol if chiral geo c}
\begin{eqnarray}
\mathcal{Q}(\varrho;t)
&=&
\frac{1}{\pi}
\frac{
\varrho^{-1/2}
     }
     {
\sqrt{4\pi t^{\prime 3}}
     }
\int_{0}^{1}{d}u\,
\frac{1}{(1-u)\sqrt{\varrho-u}}
\nonumber \\
&&
\times
\int_{0}^{1}{d}v\,
\frac{D(u,v;\varrho)}{\sqrt{v(1-v)}}
e^{-D^{2}(u,v;\varrho)/4t^{\prime}}
\nonumber\\
&&
\end{eqnarray}
with the auxiliary function and variable
\begin{eqnarray}
D(u,v;\varrho):=\ln\left[\frac{\varrho-u}{vu(1-u)}\right],
\qquad
t^{\prime}:=2t.
\end{eqnarray}
The integrations over $u$ and $v$ can be performed in closed form,
\begin{eqnarray}
\mathcal{Q}(\varrho;t)=
\frac{
1
     }
     {
\sqrt{2\pi t\varrho (\varrho-1)}
     }
\exp
\left[ 
-\frac{
\left(\mathrm{arccosh}\,\varrho^{1/2}\right)^{2}
      }
      {
2t
      }
\right].
\nonumber\\
&&
\end{eqnarray}
\end{subequations}
By shifting the minimum of the resistance $\varrho$ from 1 to 0,
$\rho:=\varrho-1$, we recover the same asymptotics for
the probability distribution of $\rho$ computed by
Stone and Joanopoulos in Ref.\ \onlinecite{Stone81}
with a log-normal distribution of the nearest-neighbor hopping amplitudes
of the discrete Schr\"odinger equation
when $\rho\to0$ or $\rho\to\infty$.
The change of variable $g=1/\varrho$
yields the probability distribution 
\begin{eqnarray}
\mathcal{G}(g;t)=
\frac{
1
     }
     {
\sqrt{2\pi t(1-g)}\,g
     }
\exp
\left[ 
-\frac{
\left(\mathrm{arccosh}\,g^{-1/2}\right)^{2}
      }
      {
2t
      }
\right]
\nonumber\\
&&
\end{eqnarray}
for the conductance $g$ in 
the chiral symmetry class (\ref{eq: def chiral regime parameters}).
The same result follows from solving the diffusion (DMPK) equation
(\ref{eq: FP chiral limit})
and performing the change of variables $g=1/\cosh^2x$.\cite{Mudry00}

\subsubsection{
Standard symmetry class
              }
\label{subsubsec: Standard symmetry class c}

In the standard symmetry class (\ref{eq: strong+weak def of std limit})
insertion of the separation-of-variable ansatz
\begin{eqnarray}
Q(\varrho,\phi;t)=
\frac{1}{2\pi}
\mathcal{Q}(\varrho;t)
\end{eqnarray}
into the Fokker-Planck equation
(\ref{eq: master equation in terms of varrho and phi})
yields
\begin{subequations}
\label{eq: Abrikosov FP}
\begin{eqnarray}
\label{eq: Abrikosov FP a}
\frac{\partial \mathcal{Q}}{\partial t}
&=&
(2\varrho-1)\frac{\partial \mathcal{Q}}{\partial \varrho}
+
\varrho(\varrho-1)\frac{\partial^{2} \mathcal{Q}}{\partial \varrho^{2}}
\end{eqnarray}
with the initial condition
\begin{eqnarray}
\label{eq: Abrikosov FP b}
\mathcal{Q}(\varrho;t=0)=\delta(\varrho-1).
\end{eqnarray}
\end{subequations}
As shown by Abrikosov in Ref.\ \onlinecite{Abrikosov81},
the normalized solution to Eq.\ (\ref{eq: Abrikosov FP})
has the integral representation
\begin{eqnarray}
\mathcal{Q}(\varrho;t)
&=&
\frac{e^{-t/4}}{\sqrt{4\pi t^{3}}}\,
\int_{0}^{1}
{d}u\,
\frac{
e^{-[D_0(u;\varrho)]^2/4t}
}
{
[u(1-u)(\varrho-u)]^{1/2}
}
D_0(u;\varrho)
\nonumber \\
&=&
\sqrt{\frac{4}{\pi t^{3}}}\,
\int_{y^{\ }_{\varrho}}^{+\infty}
{d}y\,  
\frac{\,y\,
e^{-(t/4)-(y^{2}/t)}}{
\big(
\cosh^{2}y-\varrho
\big)^{1/2}
}
\label{eq: sol if std geo c}
\end{eqnarray}
with
$
y^{\ }_{\varrho}:=
-\frac{1}{2}
\ln
\left[
2\varrho\left(1-\sqrt{1-\varrho^{-1}}\right)-1
\right]
$ and
\begin{equation}
D_0(u;\varrho)=\ln\left[\frac{\varrho-u}{u(1-u)}\right].
\end{equation}
The change of variable $g=1/\varrho$
yields the probability distribution 
\begin{eqnarray}
\mathcal{G}(g;t)
&=&
\sqrt{\frac{4}{\pi t^{3}g^{3}}}\,
\int_{y^{\ }_{g}}^{+\infty}
{d}y\,  
\frac{\,y\,
e^{-(t/4)-(y^{2}/t)}}{
\big(
g\cosh^{2}y-1
\big)^{1/2}
}
\hphantom{AAA}
\label{eq: sol if std geo c bis}
\end{eqnarray}
with 
$
y^{\ }_{g}:=
-\frac{1}{2}
\ln
\left[
2g^{-1}\left(1-\sqrt{1-g}\right)-1
\right]
$
for the conductance $g$ 
in the standard symmetry class (\ref{eq: strong+weak def of std limit}).

\subsubsection{
Crossover regime
              }
\label{subsubsec: Crossover regime}

\begin{figure}
\begin{center}
\begin{flushleft}
\hspace{1cm}(a)
\end{flushleft}
\noindent
\includegraphics[width=75mm,clip]{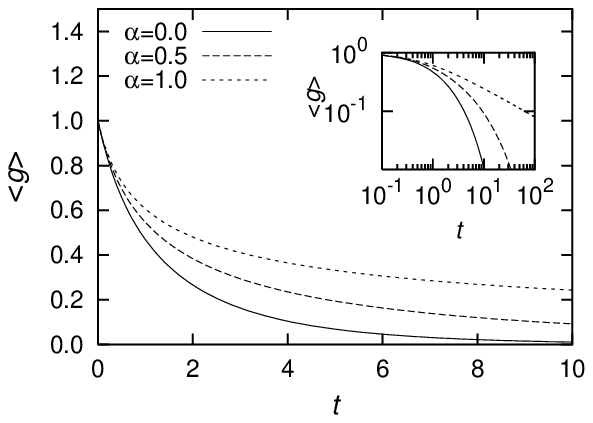}
\begin{flushleft}
\hspace{1cm}(b)
\end{flushleft}
\noindent
\includegraphics[width=75mm,clip]{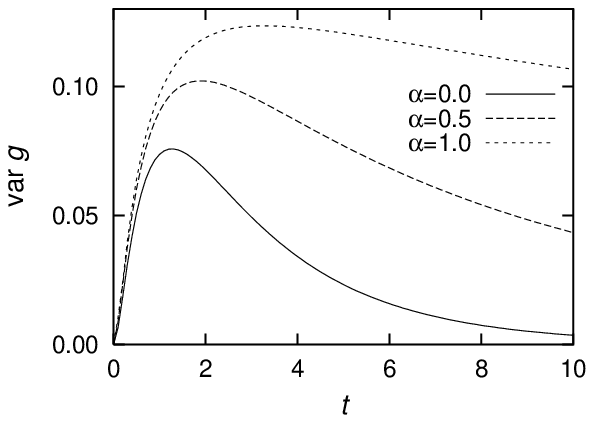}
\caption{
\label{fig: conductance}
The mean (a) and variance (b) of the conductance
computed from the probability distribution
(\ref{eq: prob dist g cross geo c})
as a function of $t=L/\ell$ for $\alpha=0,0.5,1$
and $\zeta=-1$. 
        }
\end{center}
\end{figure}

As was already the case in
Sec.\ \ref{subsubsec: Crossover regime b} 
we could not solve the Fokker-Planck equation 
(\ref{eq: master equation in terms of varrho and phi})
exactly in the crossover regime.
We thus try the separation-of-variable ansatz
\begin{eqnarray}
Q(\varrho,\phi;t)=\mathcal{Q}(\varrho;t)\Phi(\phi).
\label{eq: separation variable crossover ansatz geometry c}
\end{eqnarray}
We are again assuming that the probability distribution 
$\Phi$ is $t$ independent
and we made the specific choice for 
$\Phi$ given by Eq.\ (\ref{eq: sol cross if geo c}).
Insertion of
Eq.\ (\ref{eq: separation variable crossover ansatz geometry c})
into the Fokker-Planck equation 
(\ref{eq: master equation in terms of varrho and phi})
followed by an integration over the phase $\phi$ yields
\begin{subequations}
\label{eq: Separation FP+initial cond crossover c}
\begin{eqnarray}
\frac{\partial \mathcal{Q}}{\partial t}
&=&
-2 \zeta \alpha\mathcal{Q}
+
(2\varrho-1)(1-2\zeta \alpha )\frac{\partial \mathcal{Q}}{\partial \varrho}
\nonumber \\
&&
+
\varrho (\varrho-1)(1-\zeta \alpha)
\frac{\partial^{2} \mathcal{Q}}{\partial \varrho^{2}}
\label{eq:  Separation FP crossover c}
\end{eqnarray}
with the initial condition
\begin{eqnarray}
\mathcal{Q}(\varrho;t=0)=\delta(\varrho-1).
\label{eq:  Separation initial cond crossover c}
\end{eqnarray}
\end{subequations}
The normalized solution to 
Eq.\ (\ref{eq: Separation FP+initial cond crossover c})
has the integral representation
\begin{subequations}
\label{eq: solution cross if separation var geo c}
\begin{eqnarray}
\mathcal{Q}(\varrho;t)
&=&
\sin \left(\frac{-\pi\zeta \alpha}{1-\zeta \alpha} \right)
\frac{\varrho^{\zeta \alpha/(1-\zeta \alpha)}}
     {\sqrt{4\pi^3 t^{\prime 3}}}
\exp\left[-\frac{t^{\prime}(1+\zeta \alpha)^{2}}{4(1-\zeta \alpha)^{2}}
\right]
\nonumber \\
&&
\times
\int_{0}^{1}{d}u\,
\frac{(\frac{1}{u}-1)^{\zeta\alpha/(1-\zeta\alpha)}}
     {[u(1-u)(\varrho-u)]^{1/2}}
\nonumber \\
&&
\times
\int_{0}^{1}{d}v\,
\frac{e^{-D^{2}(u,v;\varrho)/4t^{\prime}}}
     {v^{1/2}(1-v)^{1/(1-\zeta \alpha)}}
D(u,v;\varrho)
\nonumber\\
&&
\end{eqnarray}
with the auxiliary function and variable
\begin{eqnarray}
D(u,v;\varrho):=\ln\left[\frac{\varrho-u}{vu(1-u)}\right],
\quad
t^{\prime}:=(1-\zeta \alpha)t.
\hphantom{AA}
\end{eqnarray}
\end{subequations}
As it should be, we recover
Eq.\ (\ref{eq: sol if chiral geo c})
when $\zeta\alpha=-1$
and
Eq.\ (\ref{eq: sol if std geo c})
when $\alpha=0$.
The approximate probability distribution $\mathcal{G}$ for the conductance $g$
in the crossover regime
is obtained from 
Eq.\ (\ref{eq: solution cross if separation var geo c})
through
\begin{eqnarray}
\mathcal{G}(g;t)=
g^{-2}
\mathcal{Q}(\varrho;t),
\qquad
\varrho(g)=g^{-1}.
\label{eq: prob dist g cross geo c}
\end{eqnarray}
The dependence on $t$ for the 
mean and variance of the conductance in the crossover regime
computed from 
Eq.\ (\ref{eq: prob dist g cross geo c})
are depicted in Fig.\ \ref{fig: conductance}.
In the chiral symmetry class $\zeta=-1$, $\alpha=+1$,
the mean and variance decay like $t^{-1/2}$.
For $\alpha=0.5$ and $\alpha=0$,
the mean and variance decay exponentially with $t$.

\section{
Validity of the separation ansatz for the crossover regime
        }
\label{sec: Validity of the separation ansatz for the crossover regime}

We have constructed an approximate solution to the 
Fokker-Planck equations
(\ref{eq: master equation in terms of R and phi})
and
(\ref{eq: master equation in terms of varrho and phi})
in the crossover regime (\ref{eq: crossover regime weak if a})
by assuming the separation of the 
squared magnitude $R$ from the phase $\phi$ of the reflection coefficient $r$.
In this section we shall investigate the accuracy of this approximation
to the crossover regime (\ref{eq: crossover regime weak if a})
for very large values of the length $L$ of the
disordered quantum wire, i.e., $L$ much larger than the localization
length $\xi$, which will be defined below. To this end, we shall
compare the cumulants of the radial coordinate $x$ 
computed from the separation-of-variable
ansatz with the same cumulants computed from a large-deviation ansatz
introduced by Schomerus and Titov in 
Refs.\ \onlinecite{Schomerus03a}
and 
       \onlinecite{Schomerus03b}.

\subsection{
Large $t$ limit in the crossover regime with the separation-of-variable ansatz
           }

\begin{figure}[t]
\begin{center}
\includegraphics[width=75mm,clip]{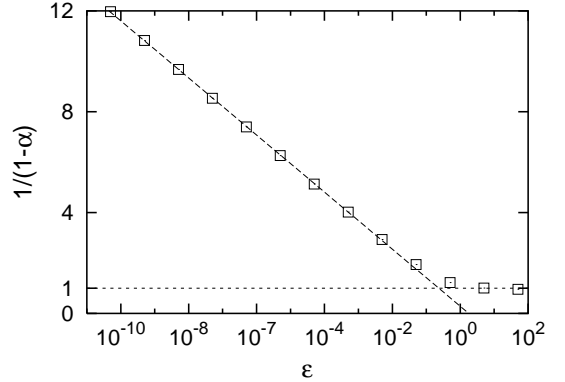}
\caption{
\label{fig: alpha}
Numerical evaluation of $1/(1-\alpha)$ as a function of $\varepsilon$
in the crossover regime (\ref{eq: crossover regime weak if a}).
The integral $\alpha$ defined in Eq.\ (\ref{eq: def alpha})
is evaluated numerically for the stationary distribution of the phase
(\ref{eq: sol cross if geo c}).
The line representing $-(1/2)\log |\varepsilon|$ is a guide to the eyes.
}
\end{center}
\end{figure}

We begin from 
Eq.\ (\ref{eq: large x limit FP x phi all derivatives to right})
without absorption and in the crossover regime.
We seek the asymptotic behavior when $t\to\infty$
of the solutions to
Eq.\ (\ref{eq: large x limit FP x phi all derivatives to right})
in the crossover regime.
We try the separation-of-variable ansatz
\begin{eqnarray}
W(x,\phi;t)=
\mathcal{X}(x;t)\Phi(\phi),
\label{eq: separation of varaibale ansatz geo c large x crossover}
\end{eqnarray}
where $\Phi$ is the stationary and normalized solution
(\ref{eq: sol cross if geo c}).
Insertion of Eq.\
(\ref{eq: separation of varaibale ansatz geo c large x crossover})
into Eq.\ (\ref{eq: large x limit FP x phi all derivatives to right})
with subsequent integration over $\phi$ yields
\begin{subequations}
\begin{eqnarray}
\frac{\partial \mathcal{X}}{\partial t}
&=&
-
\frac{1}{2}
\left(1+\zeta \alpha \right)
\frac{\partial \mathcal{X}}{\partial x}
+
\frac{1}{4}
\left(1-\zeta \alpha \right) 
\frac{\partial^{2}\mathcal{X}}{\partial x^{2}}.
\hphantom{AA}
\label{eq: FP x if large t and crossover}
\end{eqnarray}
Equation (\ref{eq: FP x if large t and crossover})
is nothing but the diffusion equation with a constant drift 
proportional to the amount of chiral symmetry breaking
$(1+\zeta\alpha)$, i.e., it is solved by
\begin{eqnarray}
\mathcal{X}(x;t)
=
\frac{
\exp
\left[
-\frac{\left[x-\left(1+\zeta \alpha\right)t/2 
\right]^{2}
     }
     {
\left(
1-\zeta \alpha\right)t}
\right]
     }
{\sqrt{\pi \left(1-\zeta \alpha\right) t}}
\label{eq: Sol x if large t and crossover}
\end{eqnarray}
\end{subequations}
normalized to the real line $x\in\mathbb{R}$.
In the approximation
(\ref{eq: Sol x if large t and crossover}),
the only nonvanishing cumulants are the first and second cumulants,
\begin{subequations}
\label{eq: cumulants x in cross geo c}
\begin{eqnarray}
\left\langle
x
\right\rangle^{\ }_c&:=&
\left\langle
x
\right\rangle
\nonumber\\
&=&
\frac{1+\zeta \alpha}{2}t,
\label{eq: first cumulant x cross geo c}
\\
\left\langle
x^2
\right\rangle^{\ }_c&:=&
\left\langle
\left(x-\frac{1+\zeta \alpha}{2}t\right)^2
\right\rangle
\nonumber\\
&=&
\frac{1-\zeta\alpha}{2}t.
\label{eq: second cumulant x cross geo c}
\end{eqnarray}
\end{subequations}
Provided $\zeta\alpha\neq-1$ 
(i.e., away from the chiral symmetry class),
the random variable $x$ becomes self-averaging in the thermodynamic limit.
It then makes sense to identify the localization length $\xi$
through 
\begin{eqnarray}
\left\langle x \right\rangle^{\ }_c=\frac{\ell}{\xi}t,
\qquad t\to\infty,
\label{eq: def xi cross geo c}
\end{eqnarray}
which, together with Eq.\ (\ref{eq: first cumulant x cross geo c}), 
gives
\begin{eqnarray}
\xi&=&
\frac{2\ell}{1+\zeta \alpha}.
\label{eq: typical localization length in cross geo c}
\end{eqnarray}
In the standard symmetry class (\ref{eq: strong+weak def of std limit})
$\alpha=0$
and  the localization length is twice the mean free path $\xi=2\times \ell$
as is well known.\cite{Beenakker97}
In the chiral symmetry class (\ref{eq: def chiral regime parameters})
$\zeta\alpha=-1$ and the localization length defined by
Eq.\ (\ref{eq: def xi cross geo c})
is not a finite number anymore but a random variable
as the right-hand side of
Eq.\ (\ref{eq: first cumulant x cross geo c}) 
vanishes whereas the right-hand side of 
Eq.\ (\ref{eq: second cumulant x cross geo c}) 
remains finite. 
For the crossover regime (\ref{eq: crossover regime weak if a}),
the constant $\alpha$ implicitly depends on
$\varepsilon$.
The asymptotic behavior of $\alpha$ for small $\varepsilon>0$ can be
evaluated as, with the probability distribution (\ref{eq: sol cross if geo c}),
\begin{eqnarray}
 \frac{1}{1-\alpha}
&=&
\frac{
\int_{-\infty}^{+\infty} {d}x\,
\frac{1}{(x^{2}+1)^{1/2}}
\int_{-\infty}^{x}{d}y\,
\frac{1}{(y^{2}+1)^{1/2}}
e^{\varepsilon (y-x)/2}
}
{
2 \int_{-\infty}^{+\infty} {d}x\, 
\frac{1}{(x^{2}+1)^{3/2}}
\int_{-\infty}^{x} {d}y\,
\frac{1}{(y^{2}+1)^{1/2}}
e^{\varepsilon (y-x)/2}
}
\nonumber \\
&\sim &
-\frac{1}{2}\ln \varepsilon,
\label{eq: alpha vs varepsilon}
\end{eqnarray}
since 
a logarithmic divergence occurs both in the $x$ and $y$ integrals 
in the numerator as we take $\varepsilon \to 0$,
while it occurs only in the $y$ integral in the denominator around $y=-\infty$.
It is then natural to identify
\begin{eqnarray}
\xi&=&
\frac{2\ell}{1+\zeta \alpha}
\qquad
(\hbox{at $\zeta=-1$ and for $\varepsilon\ll1$})
\end{eqnarray}
with the typical localization length\cite{Balents97,Eggarter78}
$(\sim |\ln|\varepsilon||\times \ell)$
when approaching the chiral symmetry class
(\ref{eq: def chiral regime parameters})
through the crossover regime 
(\ref{eq: crossover regime weak if a}).
We have also computed the dependence of $\alpha$ on $\varepsilon$ by
evaluating Eq.\ (\ref{eq: alpha vs varepsilon}) numerically.
In Fig.\ \ref{fig: alpha}
$1/(1-\alpha)$ is plotted as a function of $\varepsilon$
in the crossover regime 
(\ref{eq: crossover regime weak if a}).
We observe that 
$1/(1-\alpha)$ 
asymptotically behaves as 
$-(1/2) \ln|\varepsilon|$
in the vicinity ($\varepsilon<0.1$) of the chiral symmetry class
(\ref{eq: def chiral regime parameters}).

\subsection{
Large $t$ limit in the crossover regime with the 
large-deviation ansatz
           }

Schomerus and Titov
in Refs.\
\onlinecite{Schomerus03a}
and
\onlinecite{Schomerus03b}
devised a systematic method to compute
all the cumulants of the conductance and the LDOS
to leading order in $\xi/L\,(\ll1)$.
In practice, they carried out explicitly the computation of the
first four cumulants.

To draw a connection to their work we note that
whereas the parametrization $x=\mathrm{arctanh}\sqrt{R}$
of the magnitude of the reflection coefficient
is the natural one from a geometrical point of view
according to Eqs.\ 
(\ref{eq: polar decomposition S+M}), 
(\ref{eq: FP std limit}),
and 
(\ref{eq: FP chiral limit}), 
it is
\begin{subequations}
\begin{eqnarray}
1\ll u:=2x\sim -\ln g\sim-\ln\nu
\end{eqnarray}
that behaves statistically as the logarithm of the conductance $g$
(or the logarithm of the LDOS $\nu$ according to the Borland conjecture,
see Ref.\ \onlinecite{Borland63}) 
in the regime defined by the condition 
\begin{eqnarray}
0\leq g\ll1.
\end{eqnarray}
\end{subequations}
We denote by $U(u,\phi;L)$ the joint probability distribution
of $u$ and $\phi$, and introduce $s\equiv2L/\ell$.
We infer from 
Eq.\ (\ref{eq: large x limit FP x phi all derivatives to right}) 
that $U(u,\phi;L)$ obeys the Fokker-Planck equation
\begin{widetext}
\begin{subequations}
\label{eq: large u limit FP u phi all derivatives to right}
\begin{eqnarray}
\frac{\partial U}{\partial s}
&=&
\mathcal{L}^{\ }_{\phi} U
+
\left(
-\frac{1}{2}
+\frac{3}{2}\zeta
 \cos 2\phi
\right)
\frac{\partial U}{\partial u}
+
\frac{1}{2}
\left(
1
-
\zeta\cos 2\phi 
\right)
\frac{\partial^2U}{\partial u^2}
+\zeta
\sin (2 \phi)  
\frac{\partial^2U}{\partial u\partial\phi},
\label{eq: large u limit FP u phi all derivatives to right a}
\end{eqnarray}
where
\begin{eqnarray}
\mathcal{L}^{\ }_{\phi}
&:=&
-
\zeta
\cos 2\phi\,
-
\left(
\frac{\varepsilon}{2}
+
\frac{3}{2}\zeta \sin 2 \phi  
\right)
\frac{\partial }{\partial \phi}
+
\left(
\zeta_{0}
+\frac{1}{2}
+\frac{\zeta}{2}
\cos 2\phi 
\right)
\frac{\partial^2}{\partial \phi^2}
\nonumber\\
&=&
-\frac{\varepsilon}{2} \frac{\partial}{\partial \phi}
+\zeta_{0}\frac{\partial^{2}}{\partial \phi^{2}}
+\frac{g^{\ }_{1}}{g_{+}}
 \left(\frac{\partial}{\partial\phi}\cos\phi\right)^{2}
+\frac{g^{\ }_{2}}{g_{+}}
 \left(\frac{\partial}{\partial\phi}\sin\phi\right)^{2}
\label{eq: def mathcal L phi}
\end{eqnarray}
\end{subequations}
in the large $u$ limit and in the absence of absorption.

The fact that $\mathcal{L}^{\ }_\phi$ can be written as a total derivative
plays an essential role when solving, for any given purely imaginary $z$
and any given positive integer $k$,
the eigenvalue equation
\begin{subequations}
\label{eq: large deviation ansatz}
\begin{eqnarray}
\mu^{\ }_k f^{\ }_k=
\left[
\mathcal{L}^{\ }_{\phi}
-
z
\left(
-\frac{1}{2}
+\frac{3}{2}\zeta
 \cos 2\phi
+\zeta
\sin (2 \phi)  
\frac{\partial }{\partial\phi}
\right)
+
\frac{z^2}{2}
\left(
1
-
\zeta\cos 2\phi 
\right)
\right]
f^{\ }_k
\label{eq: large deviation ansatz a}
\end{eqnarray}
for the eigenvalue
$\mu^{\ }_{k}(z)$
and the normalized eigenfunction
$f^{\ }_{k}(z,\phi)$
of $\phi$ with periodicity $\pi$ that results from insertion of
the large-deviation ansatz
\begin{eqnarray}
U(u,\phi;L)&=&
\int_{-{i}\infty}^{+{i}\infty}
\frac{{d} z}{2\pi {i}}
\sum_{k=0}^{+\infty}
e^{\mu^{\ }_{k}(z)s-z u}
f^{\ }_{k}(z,\phi)
\label{eq: large deviation ansatz b}
\end{eqnarray}
\end{subequations}
into Eq.\ (\ref{eq: large u limit FP u phi all derivatives to right}).

For large $u\gg1$, the integration over $z$ is dominated in 
Eq.\ (\ref{eq: large deviation ansatz b})
by the region of size $\sim 1/u$ close to the origin $z=0$.
Assume that for a nonvanishing $z$ of order $1/u$,
all eigenvalues $\mu^{\ }_{k}(z)$ are real valued, of descending order,
and that the largest eigenvalue $\mu^{\ }_{k=0}(z)$ is separated from
$\mu^{\ }_{k=1}(z)$ by a finite gap. 
Consequently, truncation of the summation over $k$ to the single term $k=0$
on the right-hand side of Eq.\ (\ref{eq: large deviation ansatz b})
produces an exponentially small error
\begin{eqnarray}
U(u,\phi;L)&=&
\int_{-{i}\infty}^{+{i}\infty}
\frac{{d} z}{2\pi {i}}
\left[
e^{\mu^{\ }_{0}(z)s-z u}
f^{\ }_{0}(z,\phi)
+
\mathcal{O}\left(e^{-|\mu^{\ }_{0}(z)-\mu^{\ }_{1}(z)|s}\right)
\right].
\label{eq: large deviation ansatz c}
\end{eqnarray}
\end{widetext}

Next, assume the expansions
\begin{eqnarray}
\mu^{\ }_{0}(z)=
\sum_{n=1}^{+\infty}\mu^{(n)}_0\,z^{n},
\quad
f^{\ }_{0}(z,\phi)=
\sum_{n=0}^{+\infty}f^{(n)}_0(\phi)\,z^{n},
\label{eq: expansion mu's and f's}
\end{eqnarray}
where the expansion coefficient 
$\mu^{(n)}_0$ encodes the $n$th cumulant of $u$,
\begin{eqnarray}
\left\langle
u^n
\right\rangle^{\ }_c
&=&
n!\, \mu^{(n)}_0\, s
+\mathcal{O}\left(s^0\right).
\label{eq: relating cumulants u and mun0}
\end{eqnarray}
Insertion of the expansion (\ref{eq: expansion mu's and f's})
into the Eq.\ (\ref{eq: large deviation ansatz a})
for $k=0$ can be solved iteratively for $n=0,1,\ldots$.
For $n=0$, 
Eq.\ (\ref{eq: large deviation ansatz a})
reduces to
\begin{eqnarray}
0=\mathcal{L}^{\ }_\phi f^{(0)}_0,
\label{eq: ODE order n=0}
\end{eqnarray}
whose normalized solution
is nothing but the stationary and normalized solution to
Eq.\ (\ref{eq: FP if closed at one end and no absorption}).
For $n=1$,
Eq.\ (\ref{eq: large deviation ansatz a})
reduces to
\begin{equation}
\mu^{(1)}_0f^{(0)}_0
=
\mathcal{L}_{\phi}
f^{(1)}_0
+
\left(
 \frac{1}{2}
-\frac{3}{2}\zeta
 \cos(2\phi)
-
\zeta
\sin(2 \phi)
\partial_{\phi}
\right)
f^{(0)}_0.
\label{eq: ODE order n=1}
\end{equation}
The expansion coefficient $\mu^{(1)}_0$ is obtained from integrating
Eq.\ (\ref{eq: ODE order n=1})
over $\phi$ as a result of $\mathcal{L}^{\ }_\phi$ being a total derivative,
\begin{subequations}
\label{eq: solving for 1 cumulant u}
\begin{eqnarray}
\mu^{(1)}_0=
\frac{1}{2}
\left(
1
+
\zeta\alpha^{(0)}_0
\right),
\label{eq: computation mu10}
\end{eqnarray}
where
\begin{eqnarray}
\alpha^{(0)}_0:=
\int_0^{2\pi}{d}\phi\,\cos(2\phi) f^{(0)}_0(\phi).
\end{eqnarray}
\label{eq: computation alpha00}
\end{subequations}

\noindent 
After comparison of
Eqs.\ (\ref{eq: relating cumulants u and mun0})
and   (\ref{eq: solving for 1 cumulant u})
on the one hand with 
Eqs.\ (\ref{eq: first cumulant x cross geo c})
and   (\ref{eq: def alpha})
on the other hand, we conclude that the first cumulant
of the logarithm of the conductance is computed exactly 
to leading order in $\xi/L$
within the separation-of-variable ansatz
for the crossover regime (\ref{eq: crossover regime weak if a}).
As a corollary, the crossover 
(\ref{eq: typical localization length in cross geo c})
for the typical localization length is given exactly by
the separation-of-variable ansatz.
Carrying on the iteration to compute the expansion coefficients
$\mu^{(n)}_0$, however, we find that the separation-of-variable
ansatz breaks down for $n=1,2,\ldots$
as it does not agree anymore with the large-deviation ansatz
from Schomerus and Titov
in Refs.\
\onlinecite{Schomerus03a}
and
\onlinecite{Schomerus03b}.

\begin{figure}[t] 
\begin{center}
\begin{picture}(210,160)(-210,-120)

\thicklines
\put(-190, 20){\line(1,0){190}}
\put(-190,-10){\line(1,0){190}}
\put( 0,-10){\line(0,1){ 30}}
\put(-190,-10){\line(-2,-1){10}}
\put(-190, 20){\line(-2, 1){10}}

\thinlines
\put(-135,  5){\arc   (0, 10){ 180}}
\put(-135, 15){\vector( 1, 0){  20}}
\put(-135,-5){\line  ( 1, 0){  20}}
\put(-128,  5){$r^{\ }_{\mathrm{L}}$}

\put( -65,  5){\arc   ( 0,10){-180}}
\put( -65, 15){\line  (-1, 0){  20}}
\put( -65,-5){\vector(-1, 0){  20}}
\put( -80,  5){$r^{\ }_{\mathrm{R}}$}

\put( -100,-10){\line(0,1){30}}
\put( -103,-20){$y$}
\put(-190,-30){\vector( 1,0){190}}
\put( 0,-30){\vector(-1,0){190}}
\put( -105,-40){$L$}

\put(-210,35){$\mathrm{(a)}$}

\thicklines
\put(-200, -60){\line(1,0){200}}
\put(-200,-90){\line(1,0){200}}
\put(-200,-90){\line(0,1){ 30}}
\put( 0,-90){\line(0,1){ 30}}

\thinlines
\put(-135,-75){\arc   (0, 10){ 180}}
\put(-135,-65){\vector( 1, 0){  20}}
\put(-135,-85){\line  ( 1, 0){  20}}
\put(-128,-75){$r^{\ }_{\mathrm{L}}$}

\put( -65,-75){\arc   ( 0,10){-180}}
\put( -65,-65){\line  (-1, 0){  20}}
\put( -65,-85){\vector(-1, 0){  20}}
\put( -80,-75){$r^{\ }_{\mathrm{R}}$}

\put( -100,-90){\line(0,1){30}}
\put( -103,-100){$y$}
\put( -200,-110){\vector( 1,0){200}}
\put( 0,-110){\vector(-1,0){200}}
\put( -105,-120){$L$}

\put(-210,-50){$\mathrm{(b)}$}

\end{picture}
\end{center}
\caption{
\label{fig: BC for LDOS}
(a) 
Disordered quantum wire of length $L$ closed at the right end and connected 
to a perfect lead at the left end. 
Energy levels within the wire of length $L$
are broadened beyond the mean level spacing
by the coupling to the perfect lead.
(b) 
Disordered  quantum wire of length $L$ closed at both ends.
Energy levels within the wire are broadened beyond the mean level spacing
by the absorption $\mathrm{Im}\, E=\mathcal{E}^{\prime\prime}>0$.
        }
\end{figure}
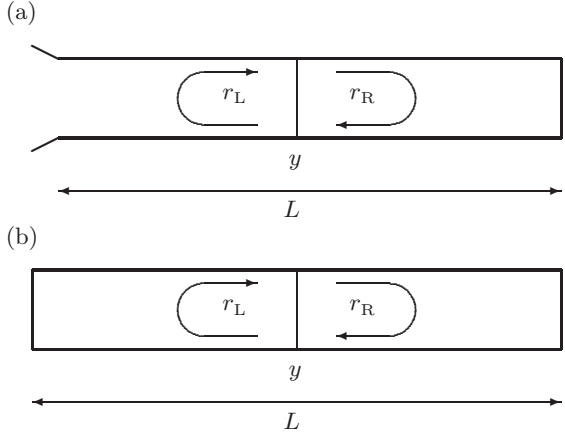

\section{
Probability distribution of the LDOS
        }
\label{sec: Probability distribution of the LDOS}

In this section,
the probability distributions
$P(\nu;L)$
and
$P(\nu;\omega)$
of the LDOS $\nu$
in the geometries of
Figs.\ \ref{fig: BC for LDOS}a 
and    \ref{fig: BC for LDOS}b, 
respectively,
are calculated based on 
the probability distribution function of the reflection 
coefficient $r=\sqrt{R}\exp({i}\phi)$
obtained in Sec.\ 
\ref{sec: Solutions to the functional renormalization group equations}.

\subsection{
Single-particle Green function and reflection coefficient
           }
\label{Single-particle Green function and reflection coefficient}

For any given realization of the disorder potentials $v^{\ }_{0,1,2}$,
the LDOS $\nu(y)$ is defined by 
\cite{Efetov97}
\begin{eqnarray}
\nu(y):=
-\frac{1}{\pi}
\mathrm{Im}\,
\mathrm{tr}\,G(y,y),
\label{eq: def meso LDOS}
\end{eqnarray}
where the matrix elements 
$G(y,z)$
in real space of the Green function are obtained from solving
\begin{eqnarray}
&&
\left[
E\!
+\!
{i}\eta
+\!
\sigma^{\ }_{3}
{i}
\frac{{d}}{{d} y}
+\!
\sum_{\mu=0}^2\sigma^{\ }_{\mu}v^{\ }_{\mu}(y)
\right]
G(y,z)=
\delta(y-z)\sigma^{\ }_{0},
\nonumber\\
&&
E=\mathcal{E}^{\prime}+{i}\mathcal{E}^{\prime\prime},
\qquad
\mathcal{E}^{\prime\prime}\geq0,
\nonumber\\
&&
(\eta:\hbox{ positive infinitesimal})
\label{eq: def Green fct}
\end{eqnarray}
with the boundary conditions associated to the geometries in
Fig.\ \ref{fig: BC for LDOS}. As long as $y$ is far away from the
ends of the disordered quantum wire, say in the middle, 
the statistical properties of
$\nu(y)$ should be independent of $y$. 
However, the statistical properties of $\nu(y)$
depend strongly on the boundary conditions
at the ends of the disordered quantum wire.\cite{Altshuler91,Schomerus02}
For completeness, we will understand under the chiral limit 
of the LDOS (\ref{eq: def meso LDOS})
the property
that the Green function (\ref{eq: def Green fct})
anticommutes with $\sigma^{\ }_1$, i.e.,
\begin{equation}
E= 
g^{\ }_0=
g^{\ }_1=0,
\label{eq: def chiral limit from Green fct}
\end{equation}
whereas we will understand under the standard limit 
of the LDOS (\ref{eq: def meso LDOS})
the conditions on the Green function (\ref{eq: def Green fct}) that
\begin{subequations}
\label{eq: def std limit from Green fct}
\begin{equation}
g^{\ }_1=g^{\ }_2
\label{eq: strong def of std limit Green}
\end{equation}
or
\begin{equation}
g^{\ }_0\gg g^{\ }_1+g^{\ }_2
\end{equation}
or
\begin{equation}
E\gg g^{\ }_1+g^{\ }_2.
\label{eq: weak def of std limit Green}
\end{equation}
\end{subequations}

Schomerus \textit{et al.}\ in Ref.\ \onlinecite{Schomerus02}
expressed the probability distribution of the LDOS in terms of 
the probability distributions of the reflection coefficient 
 $r^{\ }_{\mathrm{L}}$ 
($r^{\ }_{\mathrm{R}}$)
for outgoing plane waves reflected from the region to the left (right) of $y$
for the continuous nonrelativistic Schr\"odinger equation in the standard
symmetry class (\ref{eq: def std limit from Green fct}). 
Here we extend their analysis to 
the continuous relativistic Hamiltonian
(\ref{eq: def our Anderson localization problem a}).
A crucial observation in Ref.\ \onlinecite{Schomerus02} is that, 
for weak disorder,
the Green function can be expanded 
in terms of free scattering states within a small interval
of order of the mean free path $\ell$
since one can assume that there are no impurities
in this interval. 
The solution to Eq.\ (\ref{eq: def Green fct}) 
for $y$ and $z$ belonging to such an impurity-free interval
can be written as a linear combination of 
$e^{{i}k(+y+z)}\chi^{\ }_{+}\chi^{{T} }_{-}$,
$e^{{i}k(+y-z)}\chi^{\ }_{+}\chi^{{T} }_{+}$,
$e^{{i}k(-y+z)}\chi^{\ }_{-}\chi^{{T} }_{-}$,
$e^{{i}k(-y-z)}\chi^{\ }_{-}\chi^{{T} }_{+}$,
except for the discontinuity at $y=z$,
\begin{widetext}
\begin{eqnarray}
G(y,z)&=&
\frac{
\left(
              e^{-{i}ky}\chi^{\ }_{-}
+
r^{\ }_{\mathrm{L}}e^{+{i}ky}\chi^{\ }_{+}
\right)
\left(
e^{+{i}kz}\chi^{\ }_{-}
+
r^{\ }_{\mathrm{R}}e^{-{i}kz}\chi^{\ }_{+}
\right)^{{T} }
     }
     {
{i}\hbar v^{\ }_{\mathrm{F}}(1-r^{\ }_{\mathrm{L}}r^{\ }_{\mathrm{R}})
     }
\Theta(z-y)
\nonumber\\
&&
+
\frac{
\left(
              e^{+{i}ky}\chi^{\ }_{+}
+
r^{\ }_{\mathrm{R}}e^{-{i}ky}\chi^{\ }_{-}
\right)
\left(
e^{-{i}kz}\chi^{\ }_{+}
+
r^{\ }_{\mathrm{L}}e^{+{i}kz}\chi^{\ }_{-}
\right)^{{T} }
     }
     {
{i}\hbar v^{\ }_{\mathrm{F}}(1-r^{\ }_{\mathrm{L}}r^{\ }_{\mathrm{R}})
     }
\Theta(y-z),
\label{eq: plane wave expansion Green}
\end{eqnarray}
\end{widetext}
where
$\chi^{\ }_{+}:=(1,0)^{{T}}$,
$\chi^{\ }_{-}:=(0,1)^{{T}}$,
$\Theta(y)$ is the Heaviside function that vanishes when $y<0$ and equals
one when $y>0$, and we have momentarily 
reinstated the Fermi velocity $v^{\ }_{\mathrm{F}}$
and $\hbar$ in the relativistic dispersion relation
$\mathrm{Re}\, E\equiv\mathcal{E}^{\prime}= \hbar v^{\ }_{\mathrm{F}} k$.
The relative amplitudes among
$e^{{i}k(\pm y\pm z)}\chi^{\ }_{\pm }\chi^{{T}}_{\mp}$
as well as the factor 
$(1-r^{\ }_{\mathrm{L}}r^{\ }_{\mathrm{R}})^{-1}$
are determined by taking into account multiple scattering from the
left and right boundaries of an interval free of impurities. 

Combining 
Eqs.\ (\ref{eq: def Green fct})
and   (\ref{eq: plane wave expansion Green})
reproduces in the relativistic limit the mesoscopic
relation 
\begin{eqnarray}
\nu(y)=
\frac{1}{\pi\hbar v^{\ }_{\mathrm{F}}}
\mathrm{Re}\,
\left(
\frac{1+r^{\ }_{\mathrm{R}}r^{\ }_{\mathrm{L}}}
     {1-r^{\ }_{\mathrm{R}}r^{\ }_{\mathrm{L}}}
\right)
\label{eq: LDOS}
\end{eqnarray}
between the LDOS at $y$ and the reflection coefficients
$r^{\ }_{\mathrm{L}}$
and
$r^{\ }_{\mathrm{R}}$
from the disordered region to the left and right of $y$, respectively,
derived by Schomerus \textit{et al.}\ in Ref.\ \onlinecite{Schomerus02}
for the nonrelativistic continuous Schr\"odinger equation. 
In the nonrelativistic case,
the diagonal matrix element of the Green function in real space 
gives the microscopic local density of states.
The microscopic local density of states exhibits
$2k^{\ }_{\mathrm{F}}$ oscillations.\cite{Schomerus02}
The LDOS (\ref{eq: LDOS})
is obtained after smearing out these oscillations.
The effect of the relativistic approximation 
(\ref{eq: def our Anderson localization problem a})
is to remove
the $2k^{\ }_{\mathrm{F}}$ oscillations from the outset.
Henceforth let us measure the LDOS $\nu(y)$ in units of
$1/(\pi\hbar v^{\ }_\mathrm{F})$, the DOS of a clean wire.

The probability distribution function 
$P(\nu;L^{\ }_{\mathrm{L}},L^{\ }_{\mathrm{R}},\omega)$ 
of the LDOS $\nu(y)$ defined by
Eq.\ (\ref{eq: LDOS})
is now simply given by
\begin{widetext}
\begin{eqnarray}
P(\nu;L^{\ }_{\mathrm{L}},L^{\ }_{\mathrm{R}},\omega)
&=&
\int_0^1 {d} R^{\ }_{\mathrm{L}}
\int_0^1 {d} R^{\ }_{\mathrm{R}}
\int_0^{2\pi} {d} \phi^{\ }_{\mathrm{L}}
\int_0^{2\pi} {d} \phi^{\ }_{\mathrm{R}}\,
P^{\ }_{\mathrm{L}}
(R^{\ }_{\mathrm{L}},\phi^{\ }_{\mathrm{L}};L^{\ }_{\mathrm{L}},\omega)\,
P^{\ }_{\mathrm{R}}
(R^{\ }_{\mathrm{R}},\phi^{\ }_{\mathrm{R}};L^{\ }_{\mathrm{R}},\omega)
\nonumber\\
&&
\hphantom{
\int_0^1 {d} R^{\ }_{\mathrm{L}}
\int_0^1 {d} R^{\ }_{\mathrm{R}}
\int_0^{2\pi} {d} \phi^{\ }_{\mathrm{L}}
\int_0^{2\pi} {d} \phi^{\ }_{\mathrm{R}}
         }
\times
\delta
\left(
\nu
-
\frac{
1-R^{\ }_{\mathrm{L}}R^{\ }_{\mathrm{R}}
}
{
1+R^{\ }_{\mathrm{L}}R^{\ }_{\mathrm{R}}
-2
\sqrt{R^{\ }_{\mathrm{L}}R^{\ }_{\mathrm{R}}}
\cos\left(\phi^{\ }_{\mathrm{L}}+\phi^{\ }_{\mathrm{R}}\right)
     }
\right),
\label{eq: distribution for the LDOS}
\end{eqnarray}
\end{widetext}
where $L^{\ }_{\mathrm{L}}$ ($L^{\ }_{\mathrm{R}}$)
is the length of the segment of the disordered quantum wire
to the left (right) of the point $y$ where the LDOS is measured.
The total length of the disordered quantum wire 
is evidently given by (see Fig.\ \ref{fig: BC for LDOS})
\begin{eqnarray}
L=L^{\ }_{\mathrm{L}}+L^{\ }_{\mathrm{R}}.
\end{eqnarray}
The dependence of the probability distribution for the LDOS
on the boundary conditions enters through the dependence of
the probability distributions
$
P^{\ }_{\mathrm{L}}
(R^{\ }_{\mathrm{L}},\phi^{\ }_{\mathrm{L}};L^{\ }_{\mathrm{L}},\omega)
$
and
$
P^{\ }_{\mathrm{R}}\,
(R^{\ }_{\mathrm{R}},\phi^{\ }_{\mathrm{R}};L^{\ }_{\mathrm{R}},\omega)
$
on the boundary conditions corresponding to the geometries
of Fig.\ \ref{fig: geometries}.\cite{Altshuler91,Schomerus02}
This dependence on boundary conditions is implied by the dependence
on 
$L^{\ }_{\mathrm{L}},\omega$ 
or
$L^{\ }_{\mathrm{R}},\omega$
of the probability distribution of the two reflection coefficients
$r^{\ }_{\mathrm{L}}$ 
and
$r^{\ }_{\mathrm{R}}$,
respectively. 
When a disordered  quantum wire of finite length is closed,
one must broaden the single-particle energy levels beyond the 
mean level spacing to make sense of the energy dependence of the mesoscopic
LDOS. This is achieved here by introducing a finite absorption 
$\omega=\mathrm{Im}\, E/g^{\ }_{+}>0$, in which case the dependence on
$L^{\ }_{\mathrm{L}}\gg\ell$ 
or
$L^{\ }_{\mathrm{R}}\gg\ell$ 
becomes immaterial.
The absorption $\omega=\mathrm{Im}\, E/g^{\ }_{+}>0$ 
is switched off when a disordered  quantum wire of finite length 
$L$ is connected to a perfect lead.
We consider first the geometry depicted in
Fig.\ \ref{fig: BC for LDOS}a
and then the geometry depicted in
Fig.\ \ref{fig: BC for LDOS}b.

\begin{figure}
\begin{center}
\begin{flushleft}
(a)
\end{flushleft}
\includegraphics[width=75mm,clip]{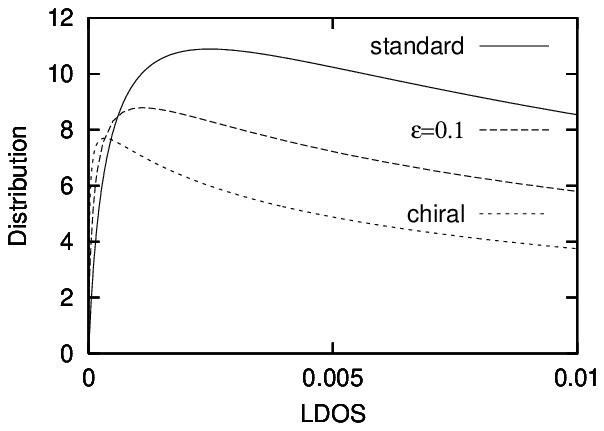} 
\begin{flushleft}
(b)
\end{flushleft}
\includegraphics[width=75mm,clip]{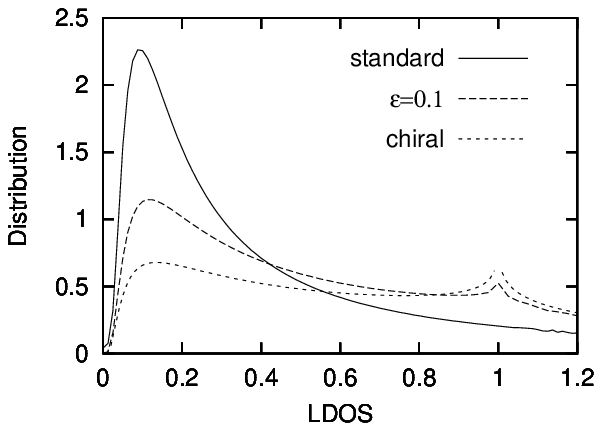}
\caption{
\label{fig: distribution of the LDOS}
(a)
Three traces for the probability distribution of the LDOS 
are plotted with
$\varepsilon=0,0.1,+\infty$ (from the bottom up at $\nu=0.01$)
so as to interpolate between the
chiral and standard symmetry classes
(\ref{eq: def chiral limit from Green fct}) 
and 
(\ref{eq: def std limit from Green fct}),
respectively.
Each trace is obtained from numerical integration 
of Eq.\ (\ref{eq: distribution for the LDOS})
for the semi-open geometry of
Fig.\ \ref{fig: BC for LDOS}a
with $L^{\ }_{\mathrm{L}}/\ell=2$.
(b)
Three traces for the probability distribution of the LDOS 
are plotted with
$\varepsilon=0,0.1,+\infty$ (from the bottom up at $\nu=0.1$)
so as to interpolate between
the chiral and standard symmetry classes
(\ref{eq: def chiral limit from Green fct}) 
and 
(\ref{eq: def std limit from Green fct}),
respectively.
Each trace is obtained from numerical integration 
of Eq.\ (\ref{eq: distribution for the LDOS})
for the closed geometry of
Fig.\ \ref{fig: BC for LDOS}b
with $\omega=1/12$.
We have verified that numerical integration of
Eq.\ (\ref{eq: distribution for the LDOS})
agrees with a representation of 
Eq.\ 
(\ref{eq: distribution for the LDOS})
in terms of elementary functions when possible.
        }
\end{center}
\end{figure}

\subsection{
Semiopen wire
           }
\label{subsec: Semi-open wire} 

To calculate the probability distribution of the LDOS
in a disordered quantum wire connected to a reservoir on the left and
closed on the right, we need
(i)
the joint probability distribution 
$P^{\ }_{\mathrm{L}}
 (R^{\ }_{\mathrm{L}},\phi^{\ }_{\mathrm{L}};L^{\ }_{\mathrm{L}})$
for the square
$R^{\ }_{\mathrm{L}}$ 
of the magnitude of the reflection coefficient $r^{\ }_{\mathrm{L}}$
and its phase $\phi^{\ }_{\mathrm{L}}$
on the segment of length $L^{\ }_{\mathrm{L}}$ to the left of the point
at which the LDOS is measured
and
(ii)
the probability distribution 
$\Phi^{\ }_{\mathrm{R}}(\phi^{\ }_{\mathrm{R}};L^{\ }_{\mathrm{R}})$
of the phase $\phi^{\ }_{\mathrm{R}}$ of the reflection coefficient
$r^{\ }_{\mathrm{R}}$
on the segment of length $L^{\ }_{\mathrm{R}}$ to the right of the point
at which the LDOS is measured.
The calculation of 
$\Phi^{\ }_{\mathrm{R}}(\phi^{\ }_{\mathrm{R}};L^{\ }_{\mathrm{R}})$
is outlined in Sec.\
\ref{subsec: disordered wire closed on the right-hand side without absorption}.
The calculation of
$P^{\ }_{\mathrm{L}}
 (R^{\ }_{\mathrm{L}},\phi^{\ }_{\mathrm{L}};L^{\ }_{\mathrm{L}})$
is outlined in Sec.\
\ref{sec: disordered wire opened at both ends without absorption}.

\subsubsection{
Chiral and standard symmetry classes
              }
\label{subsubsec: Chiral and standard symmetry classes}

The probability distribution of the 
LDOS in the chiral symmetry class 
(\ref{eq: def chiral limit from Green fct}) 
is (see the Appendix \ref{app sec: proof})
\begin{subequations}
\label{eq: log normal distribution LDOS if ch in geo semi-open}
\begin{eqnarray}
P(\nu;L^{\ }_{\mathrm{L}})&=&
\frac{
1
     }
     {
\sqrt{8 \pi L^{\ }_{\mathrm{L}}/\ell}
     }
\frac{1}{\nu}
\exp
\left(
-\frac{
\ell
      }
      {
8L^{\ }_{\mathrm{L}}
      }
\ln^{2}\nu 
\right).
\hphantom{AA}
\label{eq: prob dis LDOS chiral semi-open}
\end{eqnarray}
This is the probability distribution of a log-normal distributed
random variable with
\begin{eqnarray}
\left\langle
\ln\nu
\right\rangle^{\ }_{L^{\ }_{\mathrm{L}}}&:=&
\int_0^\infty {d}\nu\,
P(\nu;L^{\ }_{\mathrm{L}})\,
\ln\nu
\nonumber\\
&=&
0,
\\
\left\langle
(\ln\nu)^2
\right\rangle^{\ }_{L^{\ }_{\mathrm{L}}}&:=&
\int_0^\infty {d}\nu\,
P(\nu;L^{\ }_{\mathrm{L}})\,
(\ln\nu)^2\,
\nonumber\\
&=&
\frac{
4 L^{\ }_{\mathrm{L}}
     }
     {
\ell
     }.
\end{eqnarray}
\end{subequations}
The average LDOS is
\begin{equation}
\langle\nu\rangle^{\ }_{L^{\ }_{\mathrm{L}}}=
e^{2L_\mathrm{L}/\ell},
\end{equation}
which grows exponentially with $L_\mathrm{L}$.
For comparison, the probability distribution of the 
LDOS in the standard symmetry class 
(\ref{eq: def std limit from Green fct}), 
which was obtained in 
Refs.\ \onlinecite{Altshuler91}
and    \onlinecite{Schomerus02}, 
is also log-normal,
\begin{subequations}
\label{eq: log normal distribution LDOS if std in geo semi-open}
\begin{eqnarray}
&&
P(\nu;L^{\ }_{\mathrm{L}})=
\frac{
1
     }
     {
\sqrt{4\pi L^{\ }_{\mathrm{L}}/\ell}
     }
\frac{
1
     }
     {
\nu
     }
\exp
\left[
- 
\frac{
\ell
     }
     {
4L^{\ }_{\mathrm{L}}
     }
\left(
\ln\nu
+
\frac{L^{\ }_{\mathrm{L}}}{\ell}
\right)^2
\right],
\nonumber\\
&&\\
&&
\left\langle
\ln\nu
\right\rangle^{\ }_{L^{\ }_{\mathrm{L}}}=
-\frac{L^{\ }_{\mathrm{L}}}{\ell},
\\
&&
\left\langle
\left(
\ln\nu
+
\frac{L^{\ }_{\mathrm{L}}}{\ell}
\right)^2
\right\rangle^{\ }_{L^{\ }_{\mathrm{L}}}
=
\frac{2L^{\ }_{\mathrm{L}}}{\ell}.
\end{eqnarray}
\end{subequations}
Incidentally, the average LDOS is not affected by the disorder
in the standard symmetry class, 
$\langle\nu\rangle^{\ }_{L^{\ }_{\mathrm{L}}}=1$.
The probability distributions for the LDOS
in the chiral and standard symmetry classes
are depicted in
Fig.\ \ref{fig: distribution of the LDOS}a.
In the thermodynamic limit $L^{\ }_{\mathrm{L}}\to\infty$,
the logarithm of the LDOS is self-averaging in the standard symmetry class
whereas this is not the case in the chiral symmetry class.
This difference could have been anticipated from the identification
$\ln\nu\sim -2x$, valid when $L^{\ }_{\mathrm{L}}/\ell\gg1$,
with the radial coordinate $x$ that obeys the Fokker-Planck equations
(\ref{eq: FP std limit})
and
(\ref{eq: FP chiral limit}),
respectively.

\subsubsection{
Separation ansatz for the crossover regime
              }
\label{subsubsec: Separation ansatz for the crossover regime}

Numerical evaluation of the integrals on the right hand side of 
Eq.\ (\ref{eq: distribution for the LDOS})
is depicted in 
Fig.\ \ref{fig: distribution of the LDOS}a
for the crossover regime (\ref{eq: crossover regime weak if a})
within our approximation
which is encoded by the separation-of-variable ansatz
(\ref{eq: separation variable crossover ansatz geometry c}),
(\ref{eq: Separation FP+initial cond crossover c}),
(\ref{eq: solution cross if separation var geo c}),
and
(\ref{eq: sol cross if geo c}).

\subsection{
Closed wire
           }
\label{subsec: Closed wire}

To calculate the probability distribution of the LDOS
in a disordered quantum closed at both ends with a finite absorption $\omega$,
we need the joint probability distributions
$P^{\ }_{\mathrm{L}}
 (R^{\ }_{\mathrm{L}},\phi^{\ }_{\mathrm{L}};L^{\ }_{\mathrm{L}},\omega)$
and
$P^{\ }_{\mathrm{R}}
 (R^{\ }_{\mathrm{R}},\phi^{\ }_{\mathrm{R}};L^{\ }_{\mathrm{R}},\omega)$.
We can safely ignore the dependence on $L^{\ }_{\mathrm{L},\mathrm{R}}$
in the regime $L^{\ }_{\mathrm{R/L}}\gg\ell$
as is done in Sec.\ 
\ref{subsec: Disordered quantum wire closed on the right-hand side 
             with absorption}.

\subsubsection{
Chiral and standard symmetry classes
              }
\label{subsubsec: Chiral and standard symmetry classes bis}
\begin{widetext}
The probability distribution of the 
LDOS in the chiral symmetry class 
(\ref{eq: def chiral limit from Green fct})
is\cite{Brouwer02}
(see appendix \ref{app sec: proof})
\begin{subequations}
\label{eq: distribution LDOS if ch in geo closed}
\begin{eqnarray}
P(\nu;\omega)
&=&
\left[
\frac{1}{K^{\ }_{0}(\omega/2)}
\right]^{2}
\frac{1}{\nu}
\exp\left[
-\frac{\omega(1+\nu^{2})}{2\nu}
\right]
K^{\ }_{0}
\left(
\frac{\omega|1-\nu^{2}|}{2\nu}
\right)
\label{eq: distribution LDOS if ch in geo closed a}
\end{eqnarray}
where $K^{\ }_{n}(x)$, $n=0,1,\ldots$ are the modified Bessel functions
and has the asymptotics
\begin{eqnarray}
P(\nu;\omega)
&\propto&
\left\{
\begin{array}{ll}
\nu^{-1/2}\,
\exp\left(-\omega/\nu\right),
& 
\mbox{for }\nu \ll \omega/2, 
\\
&\\
-\ln (\omega |1-\nu|),
&  
\mbox{for } |1-\nu|\ll \frac{1}{\omega},
\\
&\\
\nu^{-3/2}\,
\exp \left(-\omega \nu\right),& \mbox{for } \nu \gg 2/\omega.
\end{array}
\right.
\label{eq: distribution LDOS if ch in geo closed b}
\end{eqnarray}
\end{subequations}
For comparison, the probability distribution of the 
LDOS in the standard symmetry class 
(\ref{eq: def std limit from Green fct}),
which was obtained in 
Refs.\ \onlinecite{Altshuler91}
and    \onlinecite{Schomerus02},
\begin{subequations}
\begin{eqnarray}
P(\nu;\omega)=
\frac{
\omega^2
     }
     {
\sqrt{2}\pi\nu^{3/2}
     }
\int^{\infty}_{(1+\nu^2)/(2\nu)}
{d}u
\frac{
\exp\left[-\omega(u-1)\right]
     }
     {
\sqrt{u-(1+\nu^2)/(2\nu)}
     }
\left[
u
K^{\ }_0\left(\omega\sqrt{u^2-1}\right)
+
\sqrt{u^2-1}
K^{\ }_1\left(\omega\sqrt{u^2-1}\right)
\right],
\end{eqnarray}
has the asymptotics
\begin{eqnarray}
P(\nu;\omega)
&\propto&
\left\{
\begin{array}{ll}
\nu^{-2}
\exp\left(-\omega/\nu\right),
&
\mbox{for } \nu \ll \omega/2,
\\
&
\\
\nu^{-1/2}\,
\exp\left(-\omega\nu\right),
&
\mbox{for } \nu \gg 2/\omega.
\end{array}
\right.
\end{eqnarray}
\end{subequations}
\end{widetext}
The probability distributions for the  LDOS
in the chiral and standard symmetry classes
are depicted in
Fig.\ \ref{fig: distribution of the LDOS}b.
Spectral weight from the tails $\nu\ll\omega/2$ and $\nu\gg2/\omega$
of the probability distribution in the standard symmetry class
is redistributed around $\nu=1$ in the chiral symmetry class.
Consequently, in the limit $\omega\to0$, the mean value of the
mesoscopic LDOS in the chiral symmetry class becomes
\begin{eqnarray}
\langle\nu\rangle^{\ }_{\omega}
&=&
\frac{1}{[K_0(\omega/2)]^2}
\int^\infty_1{d}\nu\left(1+\frac{1}{\nu^2}\right)
\nonumber \\
&&
\times
\exp\left[-\frac{\omega}{2}\left(\nu+\frac{1}{\nu}\right)\right]
K_0\left(\frac{\omega}{2}\left(\nu-\frac{1}{\nu}\right)\right)
\nonumber \\
&{\buildrel \omega\to0  \over =}&
\frac{1}{(\omega/2)[\ln(\omega/2)]^2}.
\end{eqnarray}
This result is identical to the smeared Dyson singularity of the
DOS in the chiral symmetry class,
\begin{eqnarray}
\nu^{\ }_{\mathrm{Dyson}}(\omega)&:=&
\int^\infty_{-\infty} {d}\varepsilon
\frac{2 \pi}{|\varepsilon|\, |\ln |\varepsilon/2||^{3}}
\frac{1}{\pi}\frac{\omega}{\varepsilon^2+\omega^2}
\nonumber\\
&{\buildrel \omega\to0  \over =}&
\frac{1}{(\omega/2)[\ln(\omega/2)]^2}.
\end{eqnarray}

\subsubsection{
Separation ansatz for the crossover regime
              }
\label{subsubsec: Separation ansatz for the crossover regime bis}

Numerical evaluation of the integrals on the right hand side of 
Eq.\ (\ref{eq: distribution for the LDOS})
is depicted in 
Fig.\ \ref{fig: distribution of the LDOS}b
for the crossover regime (\ref{eq: crossover regime weak if a})
within our approximation
which is encoded by the separation-of-variable ansatz
(\ref{eq: separation-of-variable ansatz in geometry b}),
(\ref{eq: separation ansatz FP for R geometry b})
and
(\ref{eq: sol cross if geo c}).

\section{Conclusion}
\label{sec: Conclusion}

The probability distribution of the 
mesoscopic local density of states (LDOS) $\nu$ 
for a strictly one-dimensional problem of Anderson localization 
with chiral symmetry (the chiral symmetry class)
was computed in closed form for two simply connected geometries 
assuming a weak disorder. 
As is the case when the chiral symmetry is maximally broken 
(the standard symmetry class), the probability distribution
of $\nu$ strongly depends on boundary conditions.
In a semiopen geometry we found that the  probability distribution of 
$\nu$ is log-normal with
a vanishing mean as opposed to log-normal with a finite
mean in the standard symmetry class. In a closed geometry with absorption
we found that the  probability distribution of 
$\nu$ has a double-peak structure whereby the second peak turns out to be
a logarithmic singularity at $\nu=1$.
We verified that the smeared Dyson singularity is reproduced
by the mean value of the LDOS when the absorption is sufficiently small.
Furthermore, we found one exact and proposed two approximate solutions 
to the functional renormalization group equation obeyed by the 
joint probability distribution for the 
squared modulus
and the phase of the reflection amplitude of a finite and possibly dissipative
wire with semiopen and open boundary conditions, respectively,
from which we could extract the approximate crossover of
the LDOS $\nu$ and conductance $g$ between the chiral and standard
symmetry classes. 
We could show that our approximation is exact for the first cumulant 
but fails to describe higher cumulants of the logarithms of $\nu$ and $g$
in the crossover regime.

\begin{acknowledgments}
We are indebted to P.\ W.\ Brouwer for an initial collaboration 
on this project and for his generous feedback at later stages.
We would also like to thank H.\ Schomerus and M.\ Titov
for useful comments.
This work was supported in part by
Japan Society for the Promotion of Science (S.R.).
\end{acknowledgments}

\appendix

\section{Probability distribution of the LDOS in the chiral symmetry class}
\label{app sec: proof}

In the appendix we prove
Eqs.\ (\ref{eq: prob dis LDOS chiral semi-open})
and   (\ref{eq: distribution LDOS if ch in geo closed a}).
We make use of the fact that the probability distributions for
the squared modulus and phases of the reflection coefficients
$
r^{\ }_{\mathrm{L}/\mathrm{R}}=
\tanh x^{\ }_{\mathrm{L}/\mathrm{R}}
\exp({i}\phi^{\ }_{\mathrm{L}/\mathrm{R}})
$
factorize in the chiral symmetry class. We have chosen here to represent 
$
0\leq(R^{\ }_{\mathrm{L}/\mathrm{R}})^{1/2}\leq1
$
by
$
\tanh x^{\ }_{\mathrm{L}/\mathrm{R}}
$.
We thus need to compute
\begin{widetext}
\begin{subequations}
\begin{eqnarray}
P(\nu;L^{\ }_{\mathrm{L}},L^{\ }_{\mathrm{R}},\omega)
&=&
\int_0^\infty {d} x^{\ }_{\mathrm{L}}
\int_0^\infty {d} x^{\ }_{\mathrm{R}}
\int_0^{2\pi} {d} \phi^{\ }_{\mathrm{L}}
\int_0^{2\pi} {d} \phi^{\ }_{\mathrm{R}}\,
P^{\ }_{\mathrm{L}}
(x^{\ }_{\mathrm{L}},\phi^{\ }_{\mathrm{L}};L^{\ }_{\mathrm{L}},\omega)\,
P^{\ }_{\mathrm{R}}
(x^{\ }_{\mathrm{R}},\phi^{\ }_{\mathrm{R}};L^{\ }_{\mathrm{R}},\omega)
\nonumber\\
&&
\hphantom{
\int_0^\infty {d} x^{\ }_{\mathrm{L}}
\int_0^\infty {d} x^{\ }_{\mathrm{R}}
         }
\times
\delta
\left(
\nu
-
\frac{
1-\tanh^2 x^{\ }_{\mathrm{L}}\tanh^2 x^{\ }_{\mathrm{R}}
     }
     {
1+\tanh^2 x^{\ }_{\mathrm{L}}\tanh^2x^{\ }_{\mathrm{R}}
-2
\tanh x^{\ }_{\mathrm{L}}\tanh x^{\ }_{\mathrm{R}}
\cos\left(\phi^{\ }_{\mathrm{L}}+\phi^{\ }_{\mathrm{R}}\right)
     }
\right)
\label{eq: distribution for the LDOS app a}
\end{eqnarray}
where
\begin{eqnarray}
P^{\ }_{\mathrm{L}/\mathrm{R}}
(x^{\ }_{\mathrm{L}/\mathrm{R}},
 \phi^{\ }_{\mathrm{L}/\mathrm{R}};
 L^{\ }_{\mathrm{L}/\mathrm{R}},\omega)\,
&=&
\mathcal{X}^{\ }_{\mathrm{L}/\mathrm{R}}
(x^{\ }_{\mathrm{L}/\mathrm{R}};L^{\ }_{\mathrm{L}/\mathrm{R}},\omega)\,
\frac{1}{2}
\left[
\delta\left(\phi^{\ }_{\mathrm{L}/\mathrm{R}}-0  \right)
+
\delta\left(\phi^{\ }_{\mathrm{L}/\mathrm{R}}-\pi\right)
\right].
\label{eq: distribution for the LDOS app b}
\end{eqnarray}
\end{subequations}
Integration over the angles $\phi^{\ }_{\mathrm{L}/\mathrm{R}}$
yields
\begin{eqnarray}
P(\nu;L^{\ }_{\mathrm{L}},L^{\ }_{\mathrm{R}},\omega)
&=&
\int_0^\infty {d} x^{\ }_{\mathrm{L}}
\int_0^\infty {d} x^{\ }_{\mathrm{R}}\,
\mathcal{X}^{\ }_{\mathrm{L}}
(x^{\ }_{\mathrm{L}};L^{\ }_{\mathrm{L}},\omega)\,
\mathcal{X}^{\ }_{\mathrm{R}}
(x^{\ }_{\mathrm{R}};L^{\ }_{\mathrm{R}},\omega)
\nonumber\\
&&
\times
\frac{1}{2}
\left[
\delta
\left(
\nu
-
\frac{
1-\tanh x^{\ }_{\mathrm{L}}\tanh x^{\ }_{\mathrm{R}}
     }
     {
1+\tanh x^{\ }_{\mathrm{L}}\tanh x^{\ }_{\mathrm{R}}
     }
\right)
+
\delta
\left(
\nu
-
\frac{
1+\tanh x^{\ }_{\mathrm{L}}\tanh x^{\ }_{\mathrm{R}}
     }
     {
1-\tanh x^{\ }_{\mathrm{L}}\tanh x^{\ }_{\mathrm{R}}
     }
\right)
\right].
\label{eq: distribution for the LDOS app}
\end{eqnarray}

\end{widetext}

\subsection{Semiopen wire}

Equation (\ref{eq: FP chiral limit}) 
implies that we need to evaluate 
Eq.\ (\ref{eq: distribution for the LDOS app})
with 
\begin{subequations}
\begin{eqnarray}
&&
\mathcal{X}^{\ }_{\mathrm{L}}
(x^{\ }_{\mathrm{L}};L^{\ }_{\mathrm{L}})=
\frac{2}{\sqrt{2\pi L^{\ }_{\mathrm{L}}/\ell}}
\exp\left[-\frac{x^{2 }_{\mathrm{L}}}{(2L^{\ }_{\mathrm{L}}/\ell)}\right],
\\
&&
\mathcal{X}^{\ }_{\mathrm{R}}
(x^{\ }_{\mathrm{R}};L^{\ }_{\mathrm{R}})=
\delta(x^{\ }_{\mathrm{R}}-\infty).
\end{eqnarray}
\end{subequations}
It is found that
\begin{eqnarray}
P(\nu;L^{\ }_{\mathrm{L}})&=&
\frac{1}{2}\int_{-\infty}^\infty\!\!\!\! {d}x\, 
\mathcal{X}^{\ }_{\mathrm{L}}(x;L^{\ }_{\mathrm{L}})\,
\delta\biglb( \nu-\exp(2x)\bigrb).
\nonumber\\
&&
\end{eqnarray}
Integration over $x$ yields Eq.\ (\ref{eq: prob dis LDOS chiral semi-open}).

\subsection{Closed wire}

Equation (\ref{eq: solution R for closed wire with absorption (chiral)}) 
implies that we need to evaluate 
Eq.\ (\ref{eq: distribution for the LDOS app})
with 
\begin{subequations}
\begin{eqnarray}
&&
\mathcal{X}^{\ }_{\mathrm{L}}
(x^{\ }_{\mathrm{L}};\omega)=
2\mathcal{N}^{\ }_\omega
e^{-(\omega/2)\cosh2x^{\ }_{\mathrm{L}}},
\\
&&
\mathcal{X}^{\ }_{\mathrm{R}}
(x^{\ }_{\mathrm{R}};\omega)=
2\mathcal{N}^{\ }_\omega
e^{-(\omega/2)\cosh2x^{\ }_{\mathrm{R}}},
\end{eqnarray}
\end{subequations}
where
$
\mathcal{N}^{\ }_\omega=
[K^{\ }_0(\omega/2)]^{-1}
$.
It is found that
\begin{eqnarray}
P(\nu;\omega)&=&
2\left(\mathcal{N}^{\ }_\omega\right)^2
\int\nolimits_{1}^{\infty}\!\!\!\!{d}p
\int\nolimits_{1}^{\infty}\!\!\!\!{d}q\,
\frac{e^{-\omega pq}\delta\biglb( \nu-(p/q) \bigrb)}{\sqrt{(p^2-1)(q^2-1)}}.
\nonumber\\
&&
\end{eqnarray}
Integration over $p$ and $q$ yields 
Eq.\ (\ref{eq: distribution LDOS if ch in geo closed a}).


\begin{thebibliography}{ourpaper}


\bibitem{Kramer93}
For a review see
B.\ Kramer and A.\ MacKinnon,
Rep.\ Prog.\ Phys. \textbf{56}, 1469 (1993).

\bibitem{Huckestein95} 
For a review see
B.\ Huckestein, Rev.\ Mod.\ Phys.\ \textbf{67}, 357 (1995). 

\bibitem{Dyson53}  
F.\ J.\ Dyson, 
Phys.\ Rev.\ \textbf{92}, 1331 (1953).

\bibitem{Theodorou76} 
G.\ Theodorou and M.\ H.\ Cohen, 
Phys.\ Rev.\ B \textbf{13}, 4597 (1976).

\bibitem{Eggarter78} 
T.\ P.\ Eggarter and R.\ Riedinger, 
Phys.\ Rev.\ B {\bf 18}, 569 (1978).

\bibitem{Balents97}
L.\ Balents and  M.\ P.\ A.\ Fisher,
Phys.\ Rev.\ B \textbf{56}, 12970 (1997).

\bibitem{Fleishman77}
L.\ Fleishman and D.\ C.\ Licciardello, 
J.\ Phys.\ C {\bf 10}, L125 (1977).

\bibitem{Stone81}
A.\ D.\ Stone, and J.\ D.\ Joannopoulos,
Phys.\ Rev.\ B \textbf{24}, 3592 (1981).

\bibitem{Zirnbauer96}
M.\ R.\ Zirnbauer, 
J.\ Math.\ Phys.\ \textbf{37}, 4986 (1996).

\bibitem{chRMT}
T.\ Nagao and K.\ Slevin, 
J.\ Math.\ Phys.\ \textbf{34}, 2075 (1993);
J.\ J.\ M.\ Verbaarschot and I.\ Zahed, 
Phys.\ Rev.\ Lett.\ \textbf{70}, 3852 (1993); 
S.\ Hikami and A.\ Zee, Nucl.\ Phys. 
B \textbf{408}, 415 (1993);
A.\ V.\ Andreev, B.\ D.\ Simons, and N.\ Taniguchi,
Nucl.\ Phys.\ B \textbf{432}, 487 (1994).

\bibitem{Fyodorov03}
Y.\ V.\ Fyodorov, and A.\ Ossipov,
Phys.\ Rev.\ Lett.\ \textbf{92}, 084103 (2004).

\bibitem{Brouwer00}
P.\ W.\ Brouwer, C.\ Mudry, and A.\ Furusaki, 
Phys.\ Rev.\ Lett.\ \textbf{84}, 2913 (2000);

\bibitem{Titov01}
M.\ Titov, P.\ W.\ Brouwer, A.\ Furusaki, and C.\ Mudry, 
Phys.\ Rev.\ B \textbf{63}, 235318 (2001).

\bibitem{Altland01} 
A.\ Altland and R.\ Merkt, 
Nucl.\ Phys.\ B \textbf{607}, 511 (2001).

\bibitem{Gade93} 
R.\ Gade, Nucl.\ Phys.\ B \textbf{398}, 499 (1993); 
R.\ Gade and F.\ Wegner, \textit{ibid.} \textbf{360}, 213 (1991);
F.\ J.\ Wegner, Phys.\ Rev.\ B \textbf{19}, 783 (1979).

\bibitem{Motrunich02}
O.\ Motrunich, K.\ Damle, and D.\ A.\ Huse,
Phys.\ Rev.\ B \textbf{65}, 064206 (2002).

\bibitem{Horovitz02}
B.\ Horovitz and P.\ Le Doussal,
Phys.\ Rev.\ B \textbf{65}, 125323 (2002).

\bibitem{Mudry03}
C.\ Mudry, S.\ Ryu, and A.\ Furusaki,
Phys.\ Rev.\ B \textbf{67}, 064202 (2003).

\bibitem{Altshuler91}
B.\ L.\ Altshuler and V.\ E.\ Prigodin,
Sov.\ Phys.\ JETP \textbf{68}, 198 (1989).

\bibitem{Bunder01}
J.\ E.\ Bunder and R.\ H.\ McKenzie,
Nucl.\ Phys.\ B \textbf{592}, 445 (2001).

\bibitem{Schomerus02}
H.\ Schomerus, M.\ Titov, P.\ W.\ Brouwer, and C.\ W.\ J.\ Beenakker,
Phys.\ Rev.\ B \textbf{65}, 121101(R) (2002).

\bibitem{Mudry00}
C.\ Mudry, P.\ W.\ Brouwer, and A.\ Furusaki,
Phys.\ Rev.\ B \textbf{62}, 8249 (2000).

\bibitem{Brouwer03}
P.\ W.\ Brouwer, A.\ Furusaki, and C.\ Mudry,
Phys.\ Rev.\ B \textbf{67}, 014530 (2003).

\bibitem{Deych98-01}
L.\ I.\ Deych, D.\ Zaslavsky, and A.\ A.\ Lisyansky,
Phys.\ Rev.\ Lett.\ \textbf{81}, 5390 (1998);
L.\ I.\ Deych, A.\ A.\ Lisyansky, and B.\ L.\ Altshuler,
\textit{ibid.} \textbf{84}, 2678 (2000);
L.\ I.\ Deych, M.\ V.\ Erementchouk, A.\ A.\ Lisyansky, and B.\ L.\ Altshuler,
\textit{ibid.} \textbf{91}, 096601 (2003).

\bibitem{Schomerus03a}
H.\ Schomerus, M.\ Titov,
Phys.\ Rev.\ B \textbf{67}, 100201(R) (2003).

\bibitem{Schomerus03b}
H.\ Schomerus, M.\ Titov,
Euro.\ Phys.\ J.\ B \textbf{35}, 421 (2003).

\bibitem{Titov03}
M.\ Titov and H.\ Schomerus,
Phys.\ Rev.\ Lett.\ \textbf{91}, 176601 (2003).

\bibitem{Dossetti-Romero04}
V.\ Dossetti-Romero, F.\ M.\ Izrailev, A.\ A.\ Krokhin,
Phys.\ Lett.\ A \textbf{320}, 276 (2004).

\bibitem{Schomerus00}
H.\ Schomerus and C.\ W.\ J.\ Beenakker,
Phys.\ Rev.\ Lett.\ \textbf{84}, 3927 (2000).

\bibitem{Motrunich01}
O.\ Motrunich, K.\ Damle, and D.\ A.\ Huse,
Phys.\ Rev.\ B \textbf{63}, 224204 (2001).




\bibitem{Brouwer98}
P.\ W.\ Brouwer, C.\ Mudry, B.\ D.\ Simons, and A.\ Altland, 
Phys.\ Rev.\ Lett.\ \textbf{81}, 862 (1998).

\bibitem{Lee73}
P.\ A.\ Lee, T.\ M.\ Rice, and P.\ W.\ Anderson,
Phys.\ Rev.\ Lett.\ \textbf{31}, 462 (1973).

\bibitem{Hueffmann90} 
A.\ H\"uffmann, 
J.\ Phys.\ A \textbf{23}, 5733 (1990).

\bibitem{Brouwer00-nonuni}
P.\ W.\ Brouwer, C.\ Mudry, and A.\ Furusaki, 
Nucl.\ Phys.\ B \textbf{565}, 653 (2000).


\bibitem{Anderson80}
P.\ W.\ Anderson, 
D.\ J.\ Thouless, 
E.\ Abrahams,
and 
D.\ S.\ Fisher,
Phys.\ Rev.\ B \textbf{22}, 3519 (1980).

\bibitem{Mello88}
P.\ A.\ Mello, P.\ Pereyra, and N.\ Kumar, 
Ann.\ Phys.\ (N.Y.) \textbf{181}, 290 (1988).

\bibitem{Dorokhov82}
O.\ N.\ Dorokhov, 
JETP Lett.\ \textbf{36}, 318 (1982).

\bibitem{Mathur97}
H.\ Mathur,
Phys.\ Rev.\ B \textbf{56}, 15794 (1997).

\bibitem{Shapiro86}
B.\ Shapiro,
Phys.\ Rev.\ B \textbf{34}, 4394 (1986).

\bibitem{Shapiro87}
B.\ Shapiro,
Philos.\ Mag.\ B \textbf{56}, 1031 (1987).

\bibitem{Cohen88}
A.\ Cohen, Y.\ Roth, and B.\ Shapiro,
Phys.\ Rev.\ B \textbf{38}, 12125 (1988).



\bibitem{Zinnjustin89}
J.\ Zinn-Justin,
\textit{
``Quantum field theory and critical phenomena'',
       }
(Oxford Univ.\ Press, New-York 1989).

\bibitem{footnote dicrete Langevin process}
An alternative derivation of the Fokker-Planck equation
(\ref{eq: master equation in terms of R and phi}) 
is the following.
First, a discrete Langevin process
with the discrete time step ``$\delta L$''
is written down by expanding
$r^{\ }_{\delta L}$
in
Eq.\ (\ref{eq:composition law reflection amplitude})
up to second order in  $v^{\ }_{0,1,2}$.
Second, a continuous Fokker-Planck equation for
the probability distribution
\begin{eqnarray}
P(r;L):=
\left\langle
\delta
\left(
r^{\ }_L
-
r
\right)
\right\rangle^{\ }_{\delta L}
\nonumber
\end{eqnarray}
is derived from the
discrete Langevin process using the methods of appendix A3.5
of Ref.\ \onlinecite{Zinnjustin89}, say. Here,
$\left\langle\ldots\right\rangle^{\ }_{\delta L}$
denotes averaging over $v^{\ }_{0,1,2}$ restricted to the thin
slice of length $\delta L$.

\bibitem{Abrikosov81}
A.\ A.\ Abrikosov,
Solid State Commun.\ \textbf{37}, 997 (1981).

\bibitem{Melnikov81}
V.\ I.\ Melnikov,
Fiz.\ Tverd.\ Tela.\ \textbf{23}, 782 (1981)
[Sov.\ Phys.\ Solid St.\ \textbf{23}, 444 (1981)].

\bibitem{Kumar85}
N.\ Kumar,
Phys.\ Rev.\ B \textbf{31}, 5513 (1985).

\bibitem{Rammal87}
R.\ Rammal and B.\ Doucot,
J.\ Physique  \textbf{48}, 509 (1987).



\bibitem{Kappus81}
M.\ Kappus and F.\ Wegner,
Z.\ Phys.\, B: Condens.\ Matter \textbf{45}, 15 (1981).

\bibitem{Titov04}
M.\ Titov, private communication.

\bibitem{footnote-cross}
The crossover between the chiral and standard symmetry class
induced by tuning $\zeta^{\ }_0$ or $\zeta$ with the constraint that
$\varepsilon=0$ can be studied by using the solution
(\ref{eq: Titov sol}).

\bibitem{Pradhan94}
P.\ Pradhan and N.\ Kumar,
Phys.\ Rev.\ B \textbf{50}, 9644 (1994).

\bibitem{Beenakker97} 
For a review see
C.\ W.\ J.\ Beenakker, 
Rev.\ Mod.\ Phys.\ \textbf{69}, 731 (1997).


\bibitem{Borland63}
R.\ E.\ Borland,
Proc.\ R.\ Soc.\ London, Ser.\ A \textbf{274}, 529 (1963).


\bibitem{Efetov97}
K.\ B.\ Efetov, 
\textit{``Supersymmetry in Disorder and Chaos''},
(Cambridge Univ. Press, 1997).

\bibitem{Brouwer02}
P.\ W.\ Brouwer and C.\ Mudry, unpublished.

\end{thebibliography}
\end{document}